\documentclass[useAMS,usenatbib]{mn2e}

\usepackage{graphicx}
\usepackage{amsmath}
\usepackage[dvips,usenames]{color}
\def\aj{AJ}%
\def\apj{ApJ}%
\def\apjl{ApJ}%
\def\apjs{ApJS}%
\def\aap{A\&A}%
\def\mnras{MNRAS}%
\def\pasp{PASP}%
\newcommand{\dm}{\mbox{$(m\!-\!M)$}}
\newcommand{\dmo}{\mbox{$(m\!-\!M)_{0}$}}
\newcommand{\fb}{\mbox{$f_{\rm b}$}}
\newcommand{\av}{\mbox{$A_V$}}
\newcommand{\ebv}{\mbox{$E_{B\!-\!V}$}}

\newcommand{\feh}{\mbox{\rm [{\rm Fe}/{\rm H}]}}

\newcommand{\Msun}{\mbox{$M_{\odot}$}}
\newcommand{\Teff}{\mbox{$T_{\rm eff}$}}

\newcommand{\chisqmin}{\mbox{$\chi^2_{\rm min}$}}

\newcommand{\comment}[1]{}
\newcommand{\beq}{\begin{equation}}
\newcommand{\eeq}{\end{equation}}
\newcommand{\beqa}{\begin{eqnarray}}
\newcommand{\eeqa}{\end{eqnarray}}

\title[The SFH of NGC~419]
{The star formation history of the SMC star cluster NGC~419}

\author[Rubele, Kerber \& Girardi]{Stefano Rubele$^{1,2}$,
	Leandro Kerber$^{3}$ and L\'eo Girardi$^{1}$ \\
$^{1}$ Osservatorio Astronomico di Padova -- INAF,
	Vicolo dell'Osservatorio 5, I-35122 Padova, Italy \\
$^{2}$ Dipartimento di Astronomia, Universit\`a di Padova,
	Vicolo dell'Osservatorio 2, I-35122 Padova, Italy \\
$^{3}$ Laborat\'orio de Astrof\'\i sica Te\'orica e Observacional, 
Departamento de Ci\^encias Exatas e Tecnol\'ogicas, \\
Universidade Estadual de Santa Cruz, 
Rodovia Ilh\'eus-Itabuna, km. 16 -- 
Ilh\'eus, Bahia, CEP 45662-000, Brazil
}

\begin{document}

\date{Accepted 2009 December 10.  Received 2009 December 10; 
in original form 2009 November 9}

\pagerange{\pageref{firstpage}--\pageref{lastpage}} \pubyear{2009}

\maketitle

\label{firstpage}

\begin{abstract}

The rich SMC star cluster NGC~419 has recently been found to present
both a broad main sequence turn-off and a dual red clump of giants, in
the sharp colour--magnitude diagrams (CMD) derived from the High
Resolution Channel of the Advanced Camera for Surveys on board the
{\em Hubble Space Telescope}.  In this work, we apply to the NGC~419
data the classical method of star formation history (SFH) recovery via
CMD reconstruction, deriving for the first time this function for a
star cluster with multiple turn-offs. The values for the cluster
metallicity, reddening, distance and binary fraction, were varied
within the limits allowed by present observations. The global
best-fitting solution is an excellent fit to the data, reproducing all
the CMD features with striking accuracy. The corresponding star
formation rate is provided together with estimates of its random
and systematic errors. Star formation is found to last for at least
700~Myr, and to have a marked peak at the middle of this interval, for
an age of 1.5~Gyr. Our findings argue in favour of multiple star
formation episodes (or continued star formation) being at the origin
of the multiple main sequence turn-offs in Magellanic Cloud clusters
with ages around 1~Gyr. It remains to be tested whether alternative
hypotheses, such as a main sequence spread caused by rotation, could
produce similarly good fits to the data.
\end{abstract}


\section{Introduction}
\label{intro}

Back in the nineties, the {\em Hubble Space Telescope} Wide Field
Planetary Camera 2 ({\em HST}/WFPC2) opened a new era in the study of
stellar populations in the Magellanic Clouds. Starting from
\citet{Gallagher_etal96}, many authors were able to derive the
detailed star formation history of LMC and SMC fields, via the
analysis of deep CMDs reaching well below the oldest main sequence
turn-offs (MSTO). It became possible as well the accurate measurement
of the properties and structure of many populous star clusters
\citep[e.g.][]{Mighell_etal96, Elson_etal1998, Rich_etal00, RSZ01,
GKK04, Kerber_etal07}.

The higher efficiency and larger area of the Advanced Camera for
Surveys (ACS) further improved the situation. One of the latest
achievements in the field was the conclusive evidence, based on ACS
data, that many star clusters in the LMC with ages typically larger
than 1~Gyr, present double or multiple main sequence turn-offs
\citep[MMSTO;][]{Mackey_BrobyNielsen2007, Mackey_etal08, 
Milone_etal08}. This effect was, in precedence, already indicated by
ground-based deep CMDs \citep{Bertelli_etal03,Baume_etal07}, which
however were far from presenting the fine details of the ACS ones.

While there is firm observational basis for the presence of MMSTOs in
star clusters, their interpretation is far from being settled. Once it
has been demonstrated that the presence of binaries cannot mimick the
detailed shape of MMSTOs \citep[see][]{Mackey_etal08,
Goudfrooij_etal09} the most obvious interpretation is that they are
the signature of stellar populations spanning several hundreds of Myr
in age. This interpretation however poses a major challenge to the
understanding of star formation and dynamics in star clusters, since
it is not obvious how objects with relatively shallow potential wells
could retain their gas and continue forming stars for so long a
time. \citet{Bekki_Mackey09} propose a mechanism to explain the bluest
MSTOs as being due to a second event of star formation driven by the
collision with a giant molecular cloud; this hypothesis however does
only explain {\em double} MSTOs and not the continuous MMSTO
structures observed by \citet{Goudfrooij_etal09} in NGC~1846.

\citet{Bastian_deMink09} have recently advanced a completely different 
explanation for MMSTOs: the phenomenon could be caused by the effects
of rotation during the main sequence evolution of stars. For a coeval
population, the most rapid rotators would evolve to lower \Teff\ and
generate the redder turn-offs. The hypothesis is certainly
interesting, although it requires an ad hoc decrease of rotation
velocities with the stellar mass, and relatively high rotation
velocities -- nearly half the critical break-up rotation rate -- in
order to explain the observed features.  Whether these rotation
velocities can be typical of A and F stars in the Magellanic Clouds,
and cause the large spread in \Teff\ that is apparently observed, is
yet to be clarified by new observations.

Whether the MMSTO phenomenum can be explained by stellar populations
of single or multiple ages, bears very much in the interpretation of
CMDs of nearby galaxies in general. In deriving SFH of galaxies, one
assumes that their CMDs are made by the superposition of populations
of different ages, each one presenting a narrow MSTO. Were the MSTOs
of single-burst populations intrinsically broad, most works of
SFH-recovery in nearby galaxies would have to be revised to some
extent.

Given the above situation, we decided to test if the extended-SFH
hypothesis could really explain the MMSTOs in $\sim1$-Gyr old star
clusters in the Magellanic Clouds. As a first target, we choose the
rich SMC star cluster NGC~419. Two aspects make this cluster an ideal
target for our goals. First, it does not only present a MMSTO
\citep{Glatt_etal08} but also has a dual clump of red giants
\citep{Girardi_etal09}, which is the signature of stars close to the 
transition between those that form an electron-degenerate core after
H-exhaustion, and those that do not. This additional fine structure of
the CMD provides strong constraints to the cluster mean age and to the
efficiency of convective core overshooting in stars
\citep{Girardi_etal09}, probably helping to limit the family of
stellar models that can be fit in the process of SFH-recovery. Second,
the central region of NGC~419 counts with extremely accurate
photometry provided by the High Resolution Channel (HRC) of ACS. As a
bonus, the field contamination is almost negligible for this cluster.

This paper is structured as follows: Sect.~\ref{sfh} briefly describes
the SFH-recovery method and its application to the NGC~419
data. Sect.~\ref{sec_sfh} presents the results for NGC~419 and a
discussion about the random and systematics errors in the recovered
SFH, due to the statistical fluctuations and uncertainties in
reddening, distance, metallicity, binary fraction, and field
contamination. Sect.~\ref{conclu} draws the final conclusions.

\section{Preparing NGC~419 for SFH-recovery}
\label{sfh}

\subsection{Data and photometry}
\label{data}

The dataset used in this paper has been retrieved from the HST archive
(GO-10396, PI: J.S. Gallagher) and consists of a 740~arcsec$^2$ area
centered on NGC~419, observed with the ACS/HRC in the filters F555W
and F814W. The same images have already been analysed by
\citet{Glatt_etal08, Glatt_etal09} and
\citet{Girardi_etal09}. \citet[][its figure 1]{Girardi_etal09rio}
provide a false-color version of these HRC images, together with a
comparison with the much wider ACS Wide Field Channel (WFC) images of
the same cluster. There, one may notice that the HRC field samples
just the very inner part of the cluster, and that the distribution of
stars over its area is quite uniform.

The original data has been re-reduced as described in
\citet{Girardi_etal09}. Essentially, we have accessed the archival
images already processed and calibrated using the standard procedures
mentioned in \citet{Sirianni_etal05}. Aperture and Point Spread
Function (PSF) photometry with the DAOPHOT package \citep{daophot}
were performed within a 2~pix radius, and aperture corrections were
applied. The resulting CMDs are very similar to those described in
\citet{Glatt_etal08}. The PSF photometry was then preferred and used
for all subsequent applications.

\begin{figure}
\resizebox{\hsize}{!}{\includegraphics{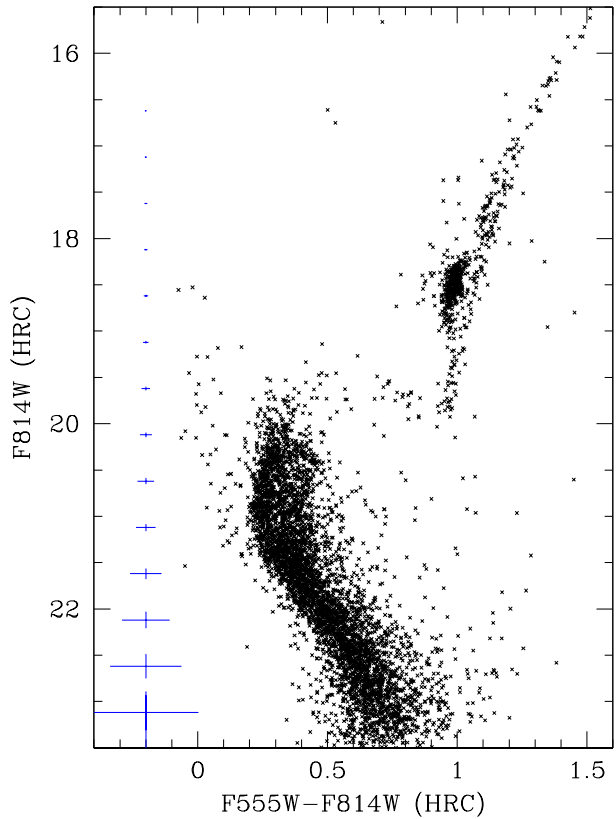}}
\caption{The CMD for NGC~419 as derived from the HRC data
centered on the cluster, after the re-reduction described in this
paper and without applying any quality cut to the photometry. The
$1\sigma$ error bars, as derived from artificial star tests (see
Sect.~\ref{sec_ast}), are drawn at the left.}
\label{fig_cmd_hrc}
\end{figure}

Fig.~\ref{fig_cmd_hrc} shows the ACS/HRC data for NGC~419, in the
F814W vs.\ F555W\,$-$\,F814W CMD. This plot will be used as a
reference to our analysis. 

\subsection{Assessing photometric errors and completeness}
\label{sec_ast}

In order to characterize the errors in the photometry and the
completeness of the sample, we have performed a series of artificial
star tests (AST) on the reduced images \citep[see
e.g.][]{Gallart_etal99,HZ01}. The procedure consists of adding stars
of known magnitude and colour at random places in the F555W and F814W
images, and redoing the photometry exactly in the same way as
before. The artificial stars are then searched in the photometric
catalogues, and when recovered the changes in their magnitudes are
stored for subsequent use.

In order to avoid the introduction of additional crowding in the
images, artificial stars are positioned at distances much higher than
their PSF width. We found that the PSF radius in on our ACS/HRC images
is of $\la7$~pix, whereas our fitting radius is of 2~pix. So, our AST
are distributed on a grid spaced by 20~pix, which is each time
randomly displaced over the image.

\begin{figure}
\resizebox{\hsize}{!}{\includegraphics{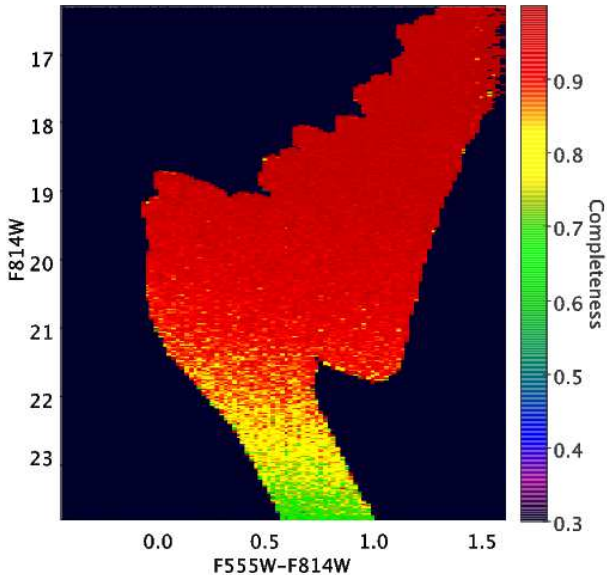}}
\caption{Completeness map, derived from the complete set of ASTs. }
\label{fig_completeness}
\end{figure}

A total of $3.4\times10^9$ ASTs were performed, covering in an almost
uniform way the CMD area of the observed stars as well as the area for
which we build the ``partial models'' to be used in the SFH analysis
(see Sect.~\ref{sec_partialmodels} below). Then, the ratio between
recovered and input stars gives origin to the completeness map of
Fig.~\ref{fig_completeness}. Note that the 90~\% completeness limit is
located at F814W\,$\sim21.7$, which is well below the position of the
MMSTOs in NGC~419. 

Another important aspect is that the stellar density is quite constant
over the HRC images, as well as the completeness. For instance, at
magnitudes 22.15, from the image center to the borders the
completeness changes from 0.87 to 0.90 for the F555W filter, and from
0.78 to 0.85 for F814W. Since this difference is very small and the
bottom part of the CMD will not be used in our analysis, we do not
apply any position-dependent completeness to the CMDs, but simply the
average values derived from all the ASTs.

\begin{figure}
\resizebox{\hsize}{!}{\includegraphics{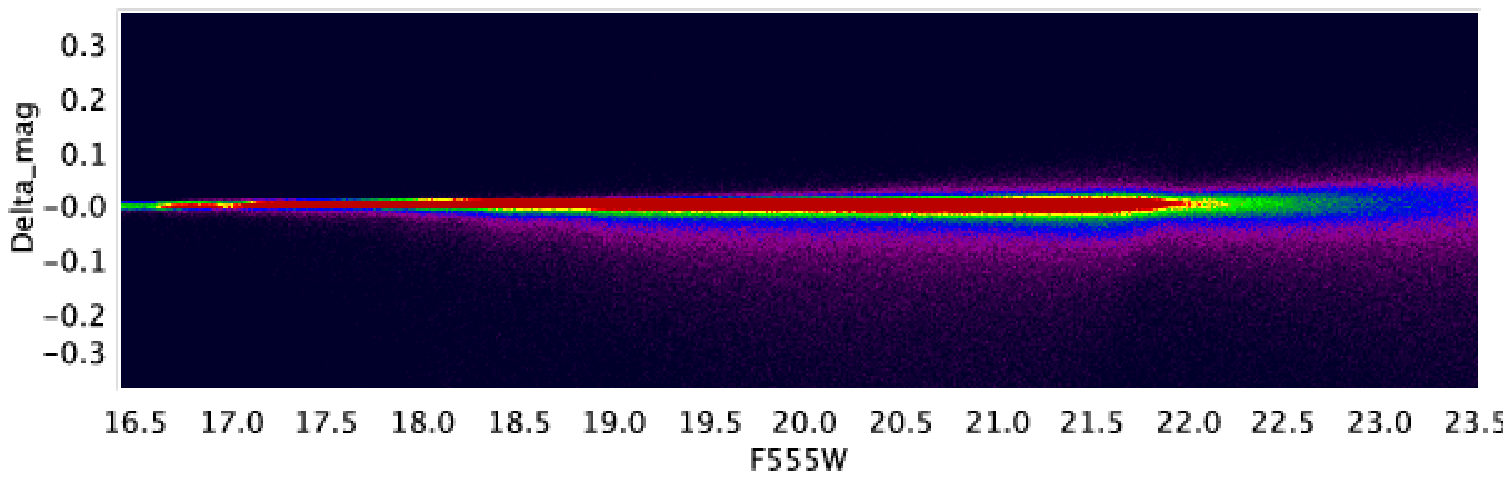}}
\resizebox{\hsize}{!}{\includegraphics{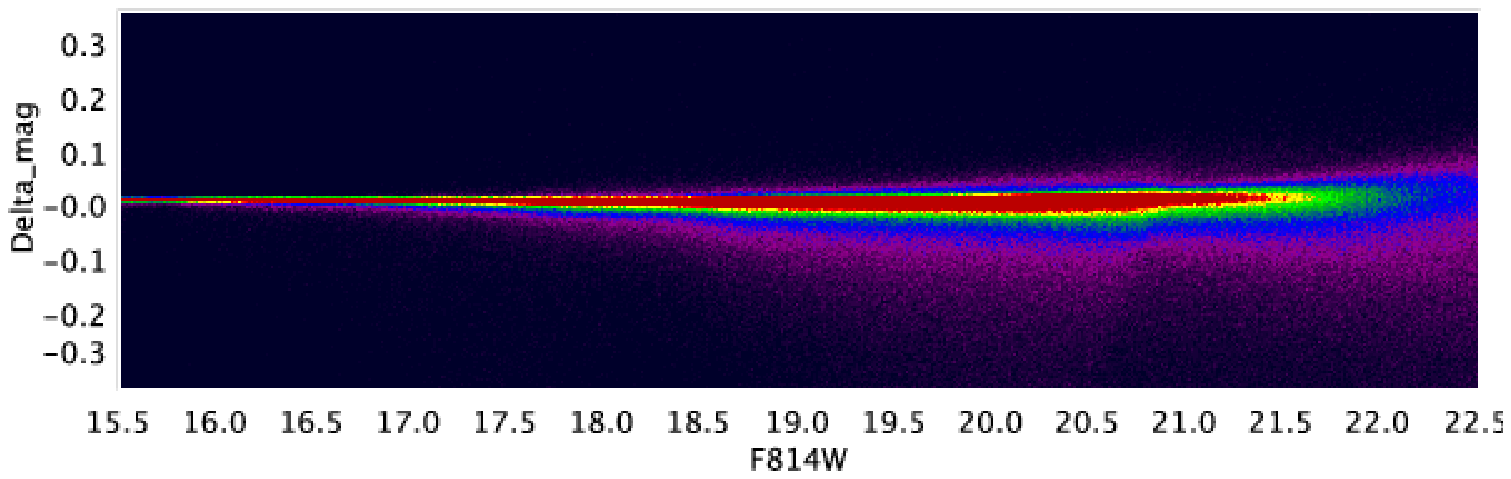}}
\caption{Map of photometric errors as a function of input F555W and
F814W, as derived from the ASTs. The errors are defined as the
difference between the recovered and input magnitudes.
}
\label{fig_photerrors}
\end{figure}

Figure~\ref{fig_photerrors} presents the differences between the
recovered and input magnitudes of ASTs, as a function of their input
magnitudes. These differences give a good handle of the photometric
errors effectively present in the data. The error distributions are
slightly asymmetric because of crowding.

\subsection{The partial models }
\label{sec_partialmodels}

The basic process of SFH-recovery is the decomposition of an observed
CMD as the sum of several ``partial models'', which represent stellar
populations in very limited intervals of age and metallicity. In our
case we will assume a constant age-metallicity relation (AMR) i.e., a
single value for the metallicity for all ages, since so far there are
no evidences for significant spreads in metallicity in star clusters
like NGC~419 \citep[e.g.][]{Mucciarelli_etal08}. Hence, each partial
model is defined as a stellar aggregate with constant star formation
over an age interval of width $\Delta\log t =0.05$~dex. This is a fine
age resolution for a work of SFH-recovery; suffice it to recall that
the age bins adopted in the SFH-recovery of nearby galaxies are
usually wider than $0.1$~dex in $\log t$ \citep[see e.g.][for some
examples]{Gallart_etal99, HZ01, Dolphin_etal03}. The age interval
covered by our partial models goes from $\log(t/{\rm yr})=8.9$ to
$9.4$, which is much wider than the interval suggested by the position
of NGC~419 MMSTOs. So, for each set of parameters, we have a total of
10 partial models, completely encompassing the age interval of
interest.

The other parameters defining a set of partial models are the $V$-band
extinction $A_V$, the distance modulus $\dmo$, the metallicity $\feh$,
and the binary fraction $\fb$; these will be varied in our analysis as
described below. Other parameters describe the area of the CMD to be
sampled, and its resolution. With basis on our Figs.~\ref{fig_cmd_hrc}
and \ref{fig_completeness}, we decide to limit the CMD area to be
analysed as: F555W$\,-\,$F814W between $-0.4$ and $1.6$, and F814W
between 16.32 and 22.0.  Within these limits, stars
are not saturated, and completeness is above 80~\%. The CMD
resolution is set to be 0.02~mag both in colour and in magnitude.

\begin{figure}
\resizebox{0.48\hsize}{!}{\includegraphics{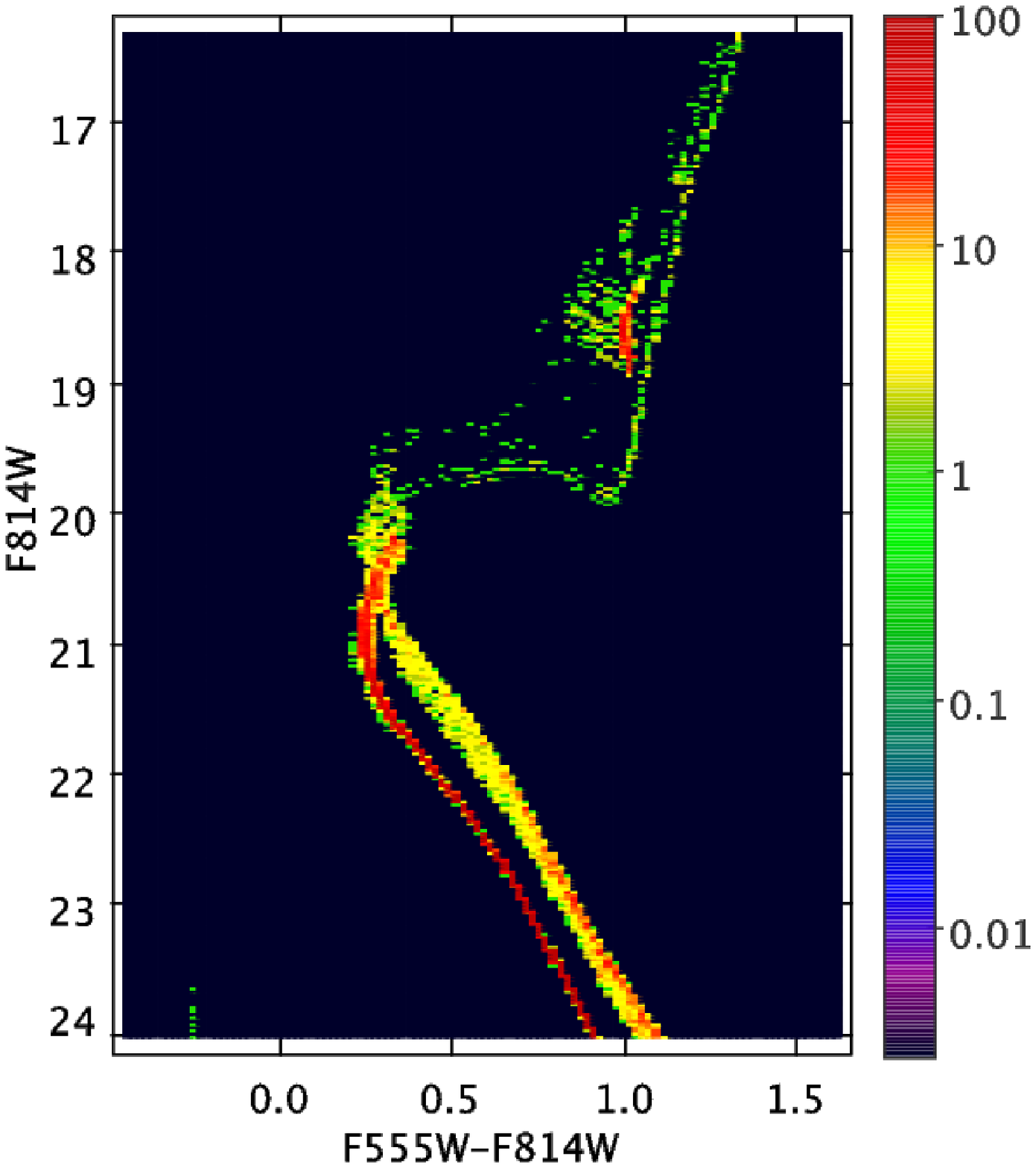}}
\resizebox{0.48\hsize}{!}{\includegraphics{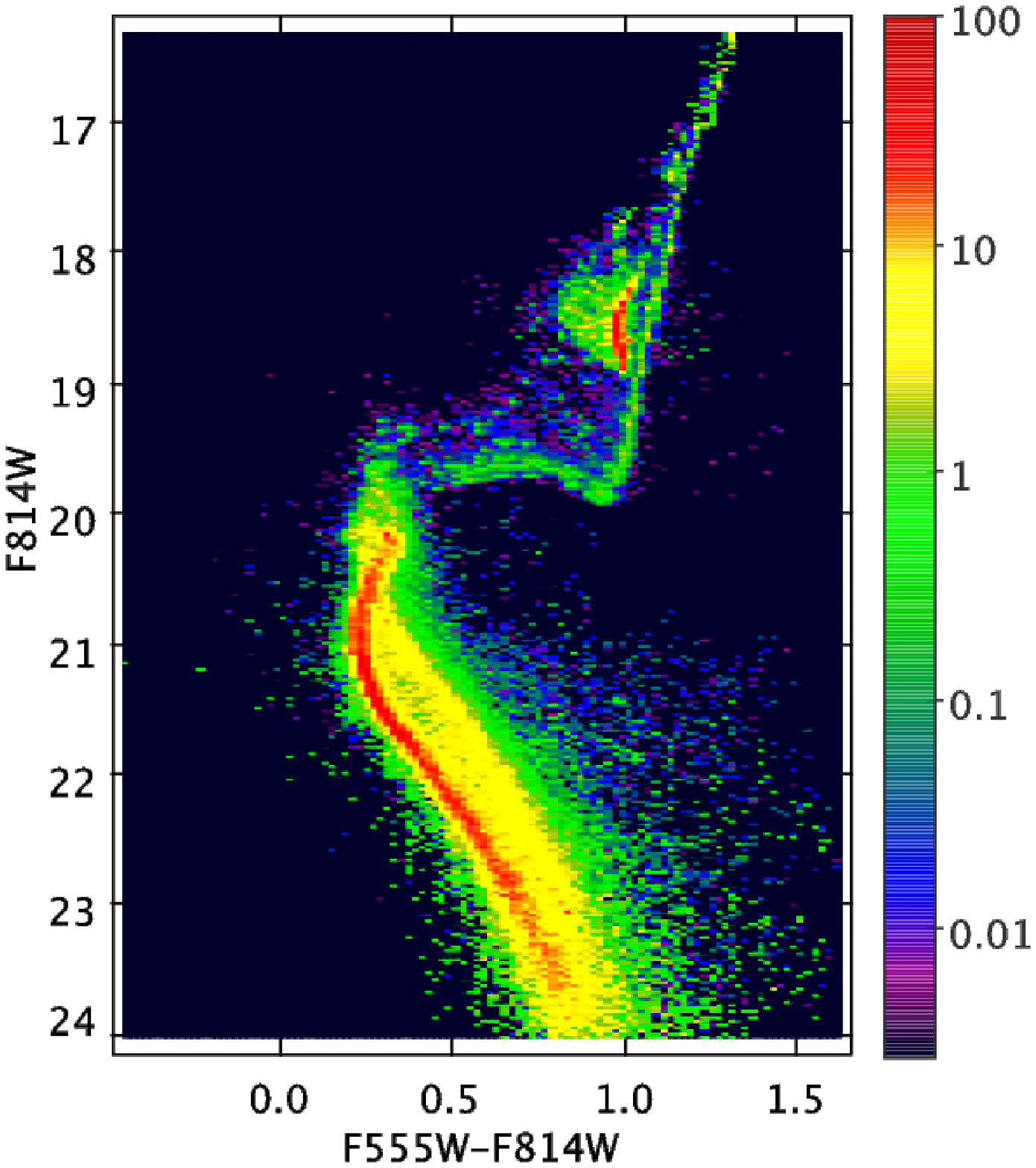}}
\caption{Hess diagram for a single partial model, before (left panel)
and after (right panel) applying the results of ASTs. The partial
model has a mean age $\log(t/{\rm yr})=9.125$, metallicity
$\feh=-0.95$, and binary fraction $\fb=0.18$. The colour scale
indicates the density of stars.}
\label{models}
\end{figure}

The partial models are simulated with the aid of the stellar
population synthesis tool TRILEGAL \citep{Girardi_etal05} in its
version 1.3, which stands on the \citet{Marigo_etal08} isochrones,
transformed to the ACS/HRC Vegamag photometry using
\citet{Girardi_etal08} bolometric corrections and extinction
coefficients. The IMF was assumed to be the \citet{Chabrier01}
lognormal one. The output catalogues from TRILEGAL are first degraded
applying the results from the ASTs -- which results in the blurring
and depopulation of the bottom part of the CMDs -- and then converted
into Hess diagrams. Fig.~\ref{models} illustrates one of such diagrams
for a partial model before and after the results of ASTs have been
applied.

\subsection{The range of cluster parameters}

In our analysis, we have tried to cover the complete possible range of
parameters for NGC~419, in particular:

\paragraph*{Distance modulus $\dmo$:} Our reference value for the SMC 
distance modulus is $\dmo = 18.89 \pm 0.14$ \citep{Harries_elal2003},
which is a value very similar to many other recent
determinations\citep[e.g.][]{Crowl_etal2001, Lah_etal2005,
North_etal2009}. In practice, however, we explored
\dmo\ values between 18.75 and 18.97.

\paragraph*{Extinction $A_V$:} The \citet{Schlegel_etal1998} 
extinction maps provide $\ebv=0.10\pm0.01$ for an area of radius equal
to $5\arcmin$ around NGC~419, which translates into
$\av=0.32\pm0.03$. A similar search in the maps from
\citet{Zaritsky_etal2002}, with a $12\arcmin$ radius, produces
$\av=0.27\pm0.28$. In practice, we explored \av\ values between 0.12
and 0.38.

\paragraph*{Metallicity \feh:} Works based on isochrone fitting  
\citep[e.g.][]{Durand_etal1984} as well as Ca~{\sc ii} triplet 
observations \citep{Kayser2009} have suggested \feh\ values similar to
$-0.7\pm0.3$ for NGC~419. Interestingly, the metal abundances of other
SMC clusters with similar ages are uncertain as well. Suffices it to
mention the recent AMR for the SMC as derived from
\citet[][their figure 14]{Parisi_etal09}, which shows SMC clusters
with ages smaller than 2~Gyr covering the complete \feh\ interval
between $-0.5$ and $-1.1$. The situation is far from clear and high
resolution spectroscopy of a handful of giants in 1-Gyr old SMC
clusters is definitely needed. In this work, we have decided to
explore the whole $-0.5<\feh<-1.1$ interval.

\paragraph*{Binary fraction \fb:} \citet{Elson_etal1998}, using the 
shape of the main sequence in a deep CMD (obtained with HST) for the
LMC cluster NGC~1818, determined a binary fraction from $\sim 0.20$ to
$\sim 0.35$, from the center to the outer parts of the cluster.  These
binary fractions refer only to the binaries with high
primary/secondary mass ratio, say of above 0.7, because these are the
only ones which separate clearly from the single-star main sequence in
CMDs. In our previous work on NGC~419
\citep{Girardi_etal09}, we have already noticed that the HRC CMD was
indicating a binary fraction of the order of 0.2, consistent with the
central region of NGC~1818. In the following, we will adopt a
reference value of $\fb\sim 0.18$; notice however that we have
explored \fb\ values from 0.10 to 0.28. The distribution of
primary/secondary mass ratios is assumed to be uniform over the
interval from 0.7 to 1.0

\paragraph*{}
Compared to the other parameters, the binary fraction is the less
important in defining the quality of a CMD fit, as will be illustrated
below. A clear advantage of exploring ranges for the other parameters
-- instead of taking single values from the literature -- is that in
this way we can partially compensate for the possible errors and
offsets in the theoretical models: Errors in the zeropoints of
bolometric corrections and on the helium content of stellar models
could be easily compensated by a change in apparent distance modulus
$\dmo+\av$. Errors in the theoretical \Teff-color relations can also
be compensated by small changes in both \av\ and \feh. Therefore, the
final best-fitting values for these quantities are to be associated to
the stellar models which we are using. Different sets of stellar
models are likely to produce slightly different best-fitting values.

\section{Recovering the Star Formation History}
\label{sec_sfh}

\subsection{Method}

To recover the SFH of NGC~419, we use an adapted version of the
pipeline that is being built to analyse data from the VISTA survey of
the Magellanic Clouds \citep[VMC; see][]{Cioni_etal08}. The method is
fully described, and tested on VMC simulated data by
\citet{Kerber_etal2009}. In brief, after building the Hess diagram for
the data and partial models, the StarFISH code \citep{HZ01,HZ04} is
used to find the linear combination of partial models that minimizes a
$\chi^2$-like statistic as defined in \citet{Dolphin02}. The solution
is characterized by the minimum $\chi^2$, $\chisqmin$, and by a set of
partial model coefficients corresponding to the several age
bins. These latter translate directly in the star formation rate as
function of age, SFR$(t)$.

\begin{figure*}
\begin{minipage}{0.32\textwidth}
\resizebox{\hsize}{!}{\includegraphics{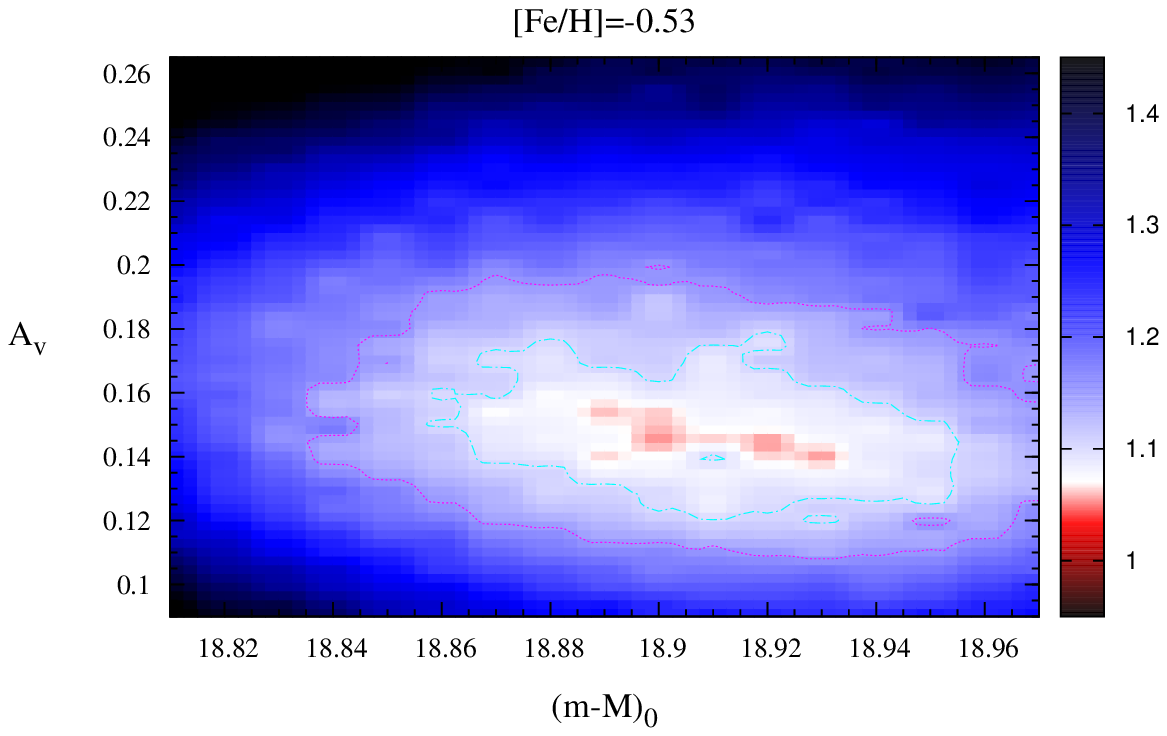}}
\end{minipage}
\begin{minipage}{0.32\textwidth}
\resizebox{\hsize}{!}{\includegraphics{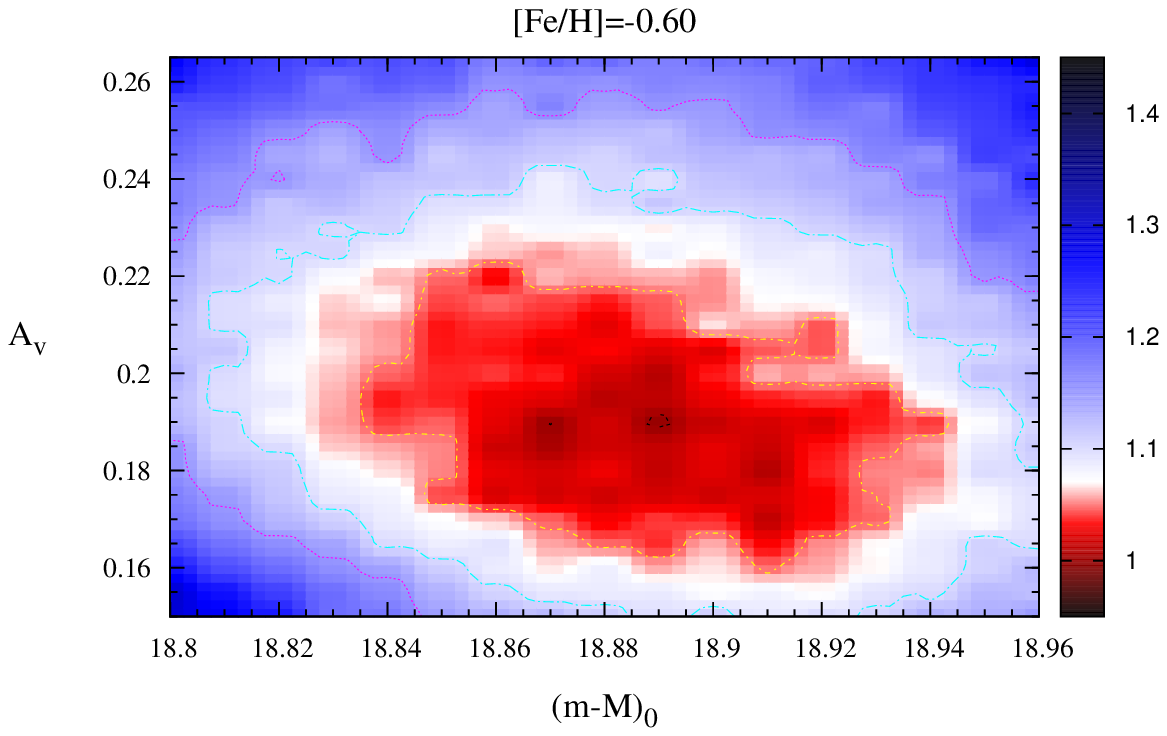}}
\end{minipage}
\begin{minipage}{0.32\textwidth}
\resizebox{\hsize}{!}{\includegraphics{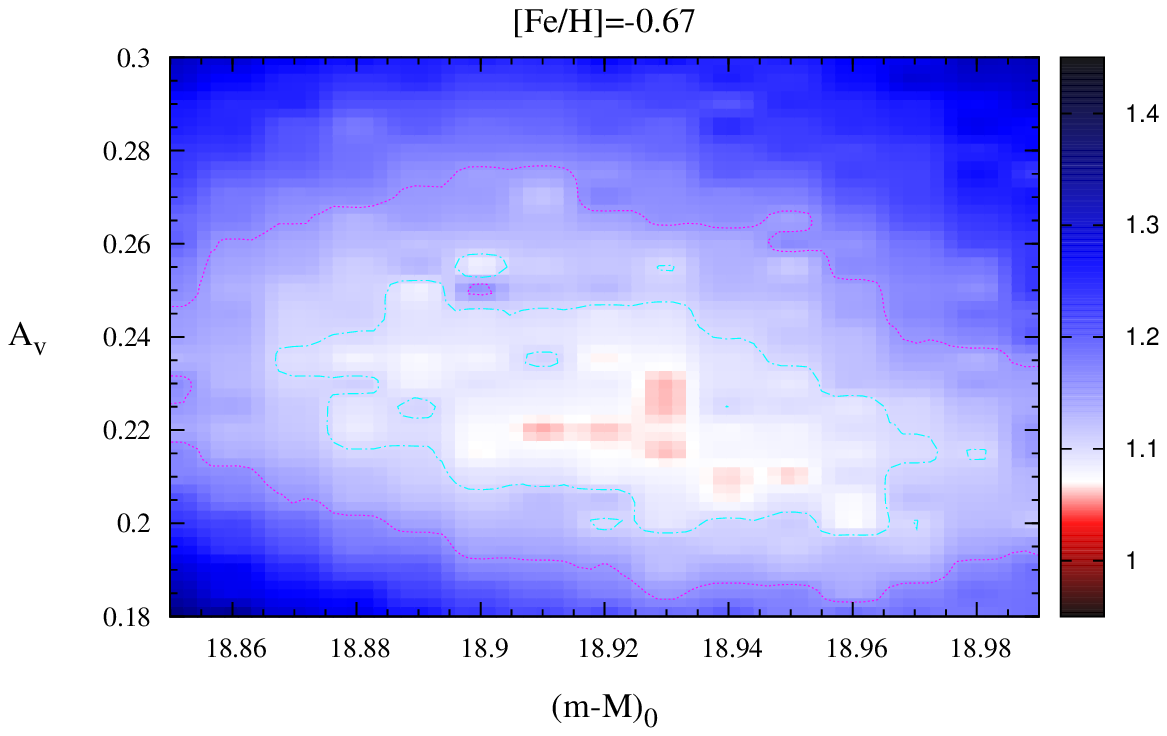}}
\end{minipage}
\begin{minipage}{0.32\textwidth}
\resizebox{\hsize}{!}{\includegraphics{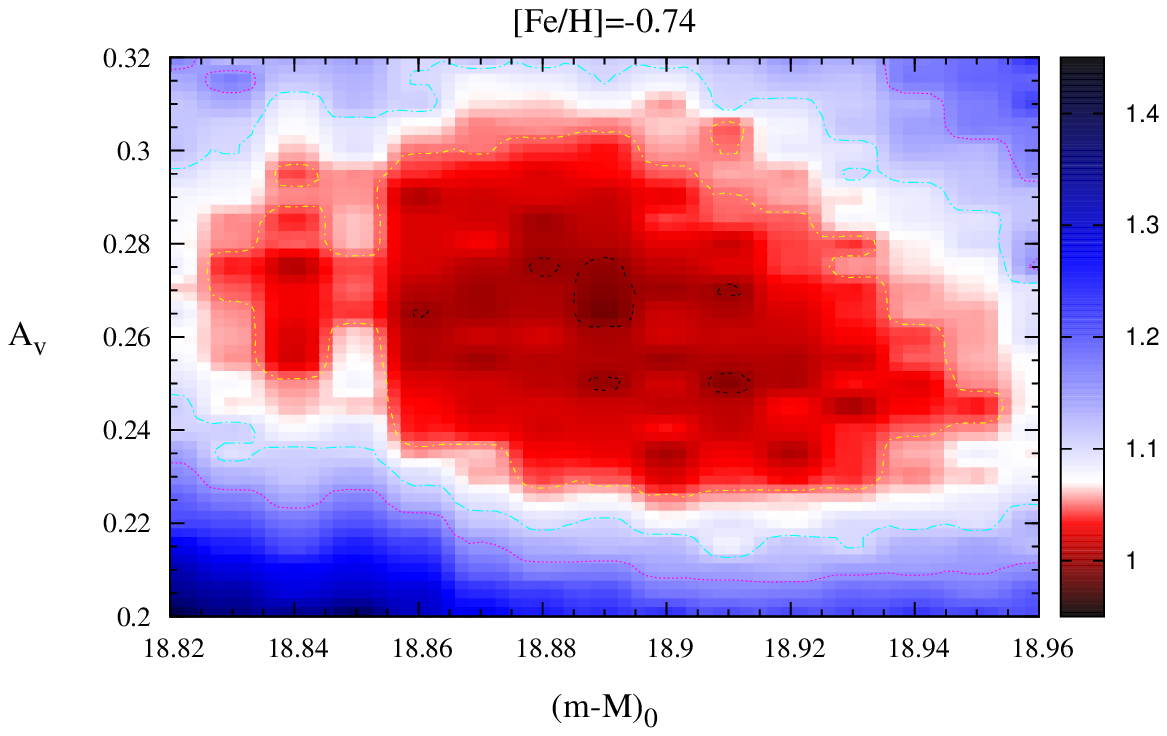}}
\end{minipage}
\begin{minipage}{0.32\textwidth}
\resizebox{\hsize}{!}{\includegraphics{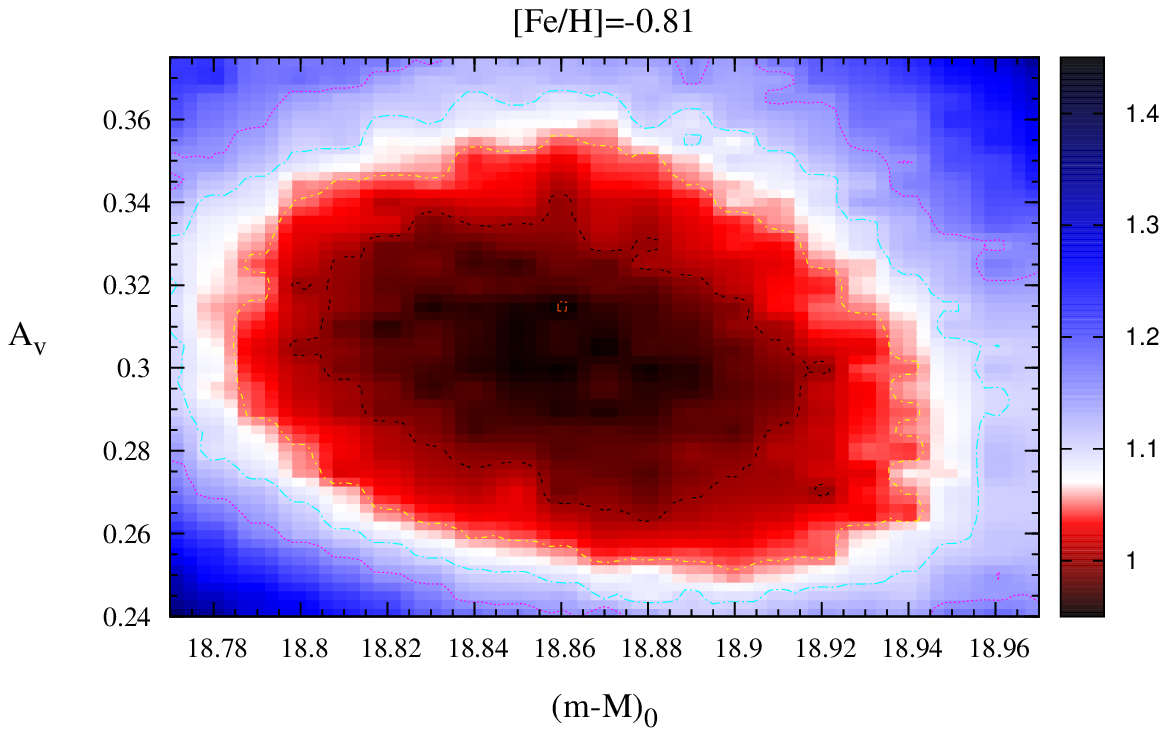}}
\end{minipage}
\begin{minipage}{0.32\textwidth}
\resizebox{\hsize}{!}{\includegraphics{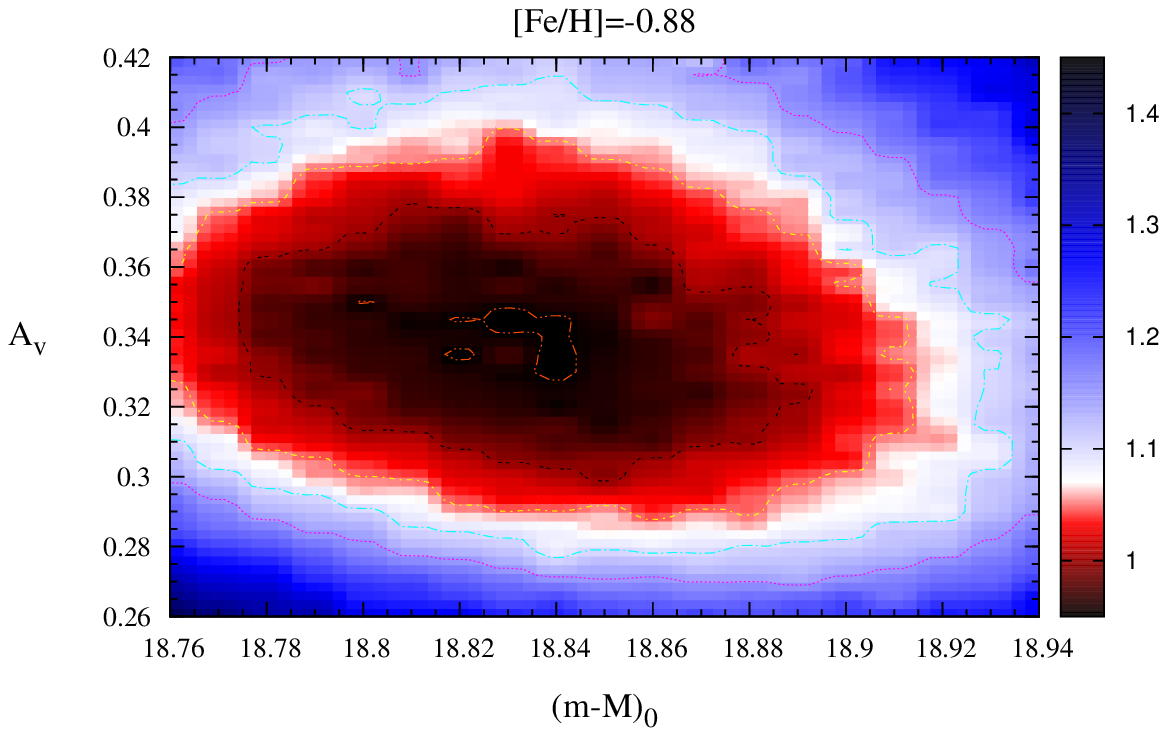}}
\end{minipage}
\begin{minipage}{0.32\textwidth}
\resizebox{\hsize}{!}{\includegraphics{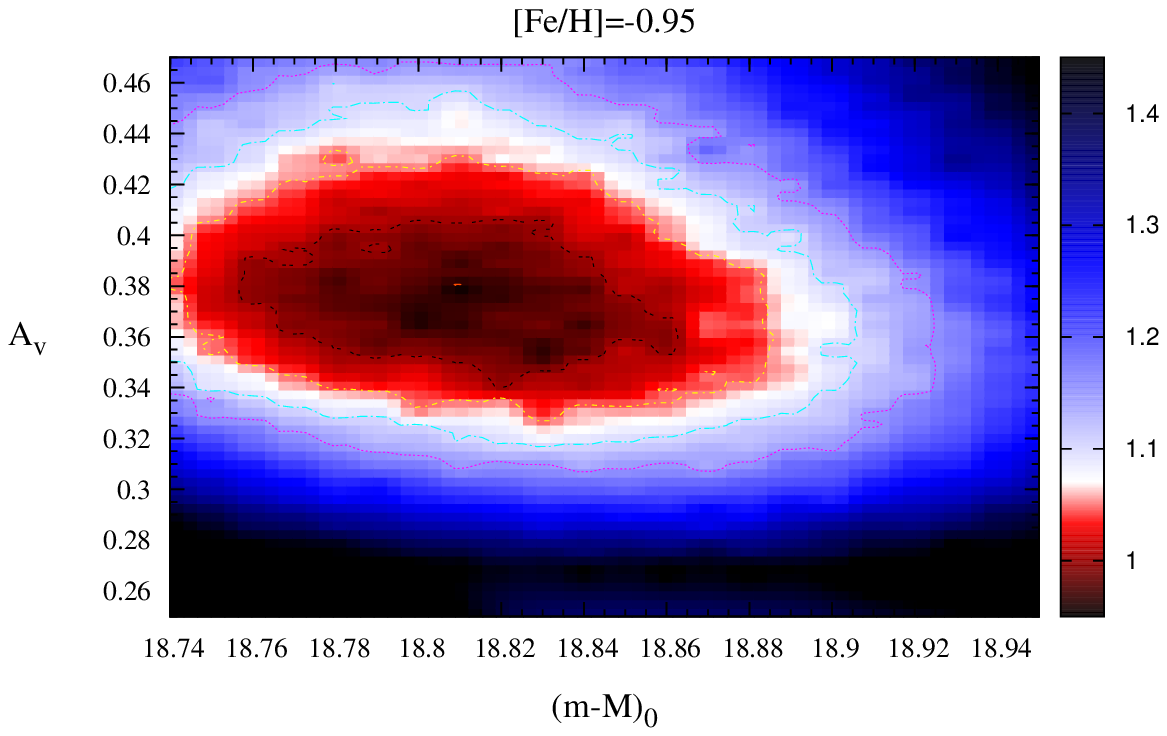}}
\end{minipage}
\begin{minipage}{0.32\textwidth}
\resizebox{\hsize}{!}{\includegraphics{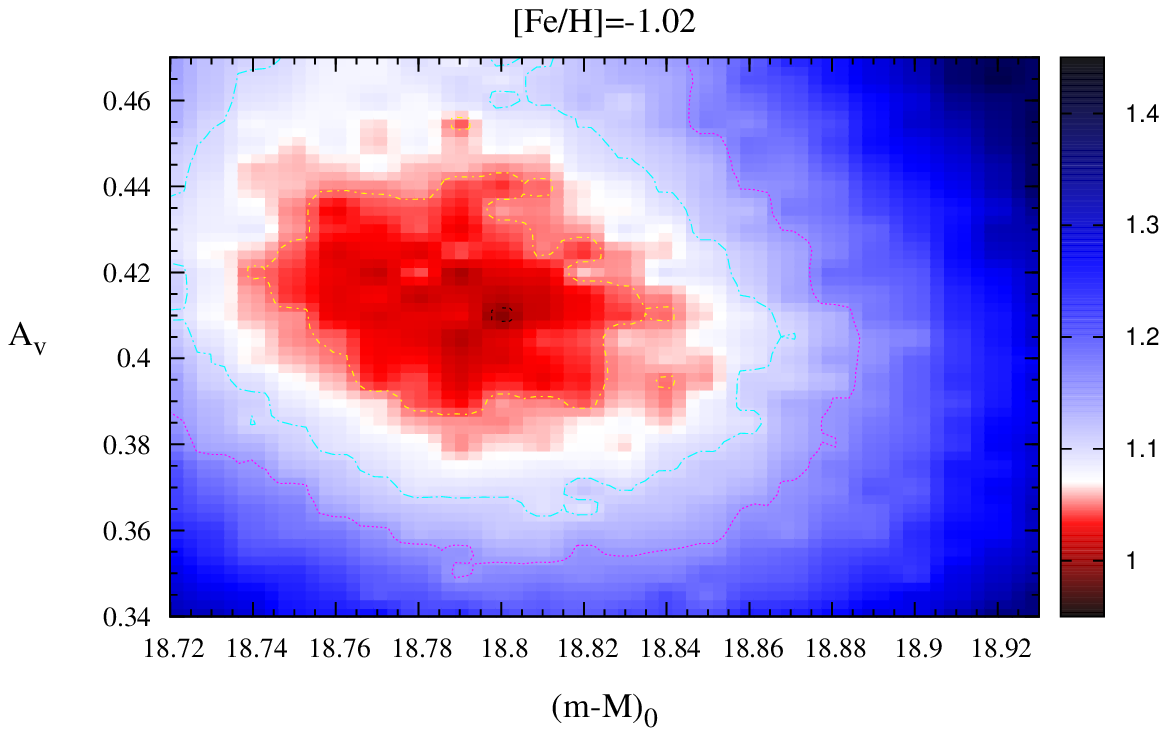}}
\end{minipage}
\begin{minipage}{0.32\textwidth}
\resizebox{\hsize}{!}{\includegraphics{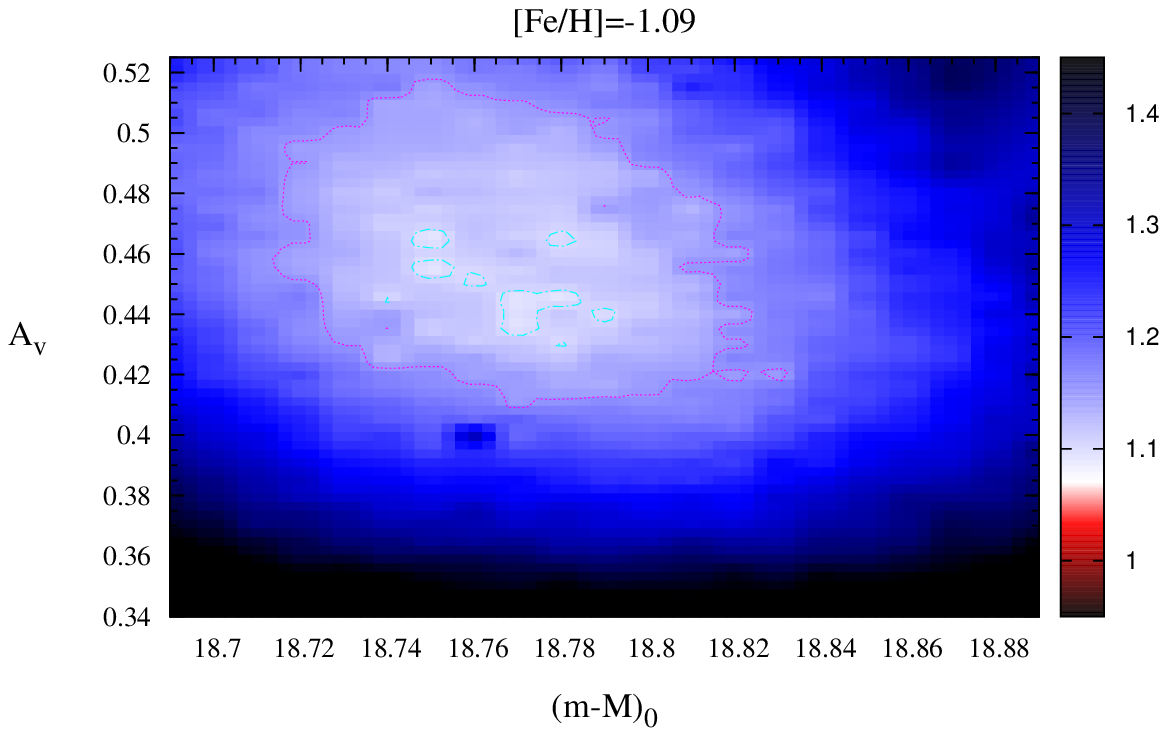}}
\end{minipage}
\caption{Maps of the $\chisqmin$ obtained during SFH-recovery, 
$\chisqmin$, as a function of \dmo\ and \av, for several \feh\
values.}
\label{chi2_map}
\end{figure*}

The SFH-recovery is repeated for each of the \dmo, \av, \feh, and \fb\
values.  In order to limit the space of parameters to be covered, the
procedure is essentially the following: for a given value of \feh\ and
\fb, we cover a significant region of the \dmo\ versus \av\ plane,
performing SFH-recovery for each point in a grid, and then building a
map of the $\chisqmin$ for the solutions. Examples of these maps are
presented in Fig.~\ref{chi2_map}. The maps are extended so that the
minimum $\chisqmin$ for a given value of $\feh$ can be clearly
identified, as well as the regions in which $\chisqmin$ increases by a
factor of about 1.5. The typical resolution of such maps is of 0.01 in
\dmo\ and 0.005 in \av.

The procedure is done for several values of \feh. $\chisqmin$ maps for
different \feh\ present relative minima in slightly different positions, so
that the final resulting maps do not cover exactly the same \av\ and
\dmo\ intervals. In any case, the relative minima are well 
delimitated in all cases, as illustrated in Fig.~\ref{prob_map}. 
The \feh\ spacing between these maps is of $\Delta\feh=0.07$~dex.

\begin{figure*}
\begin{minipage}{0.24\textwidth}
\resizebox{\hsize}{!}{\includegraphics{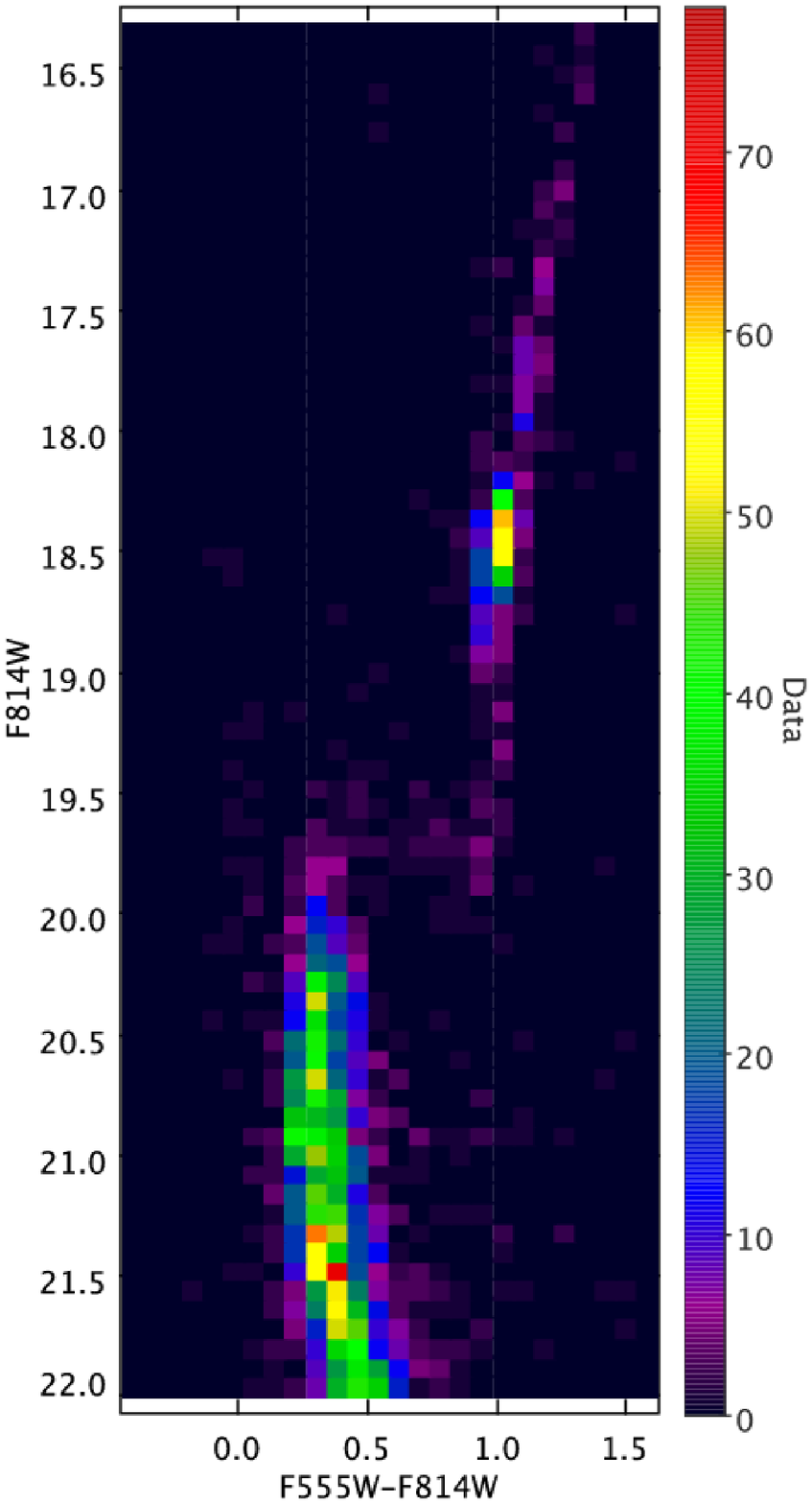}}
\end{minipage}
\begin{minipage}{0.24\textwidth}
\resizebox{\hsize}{!}{\includegraphics{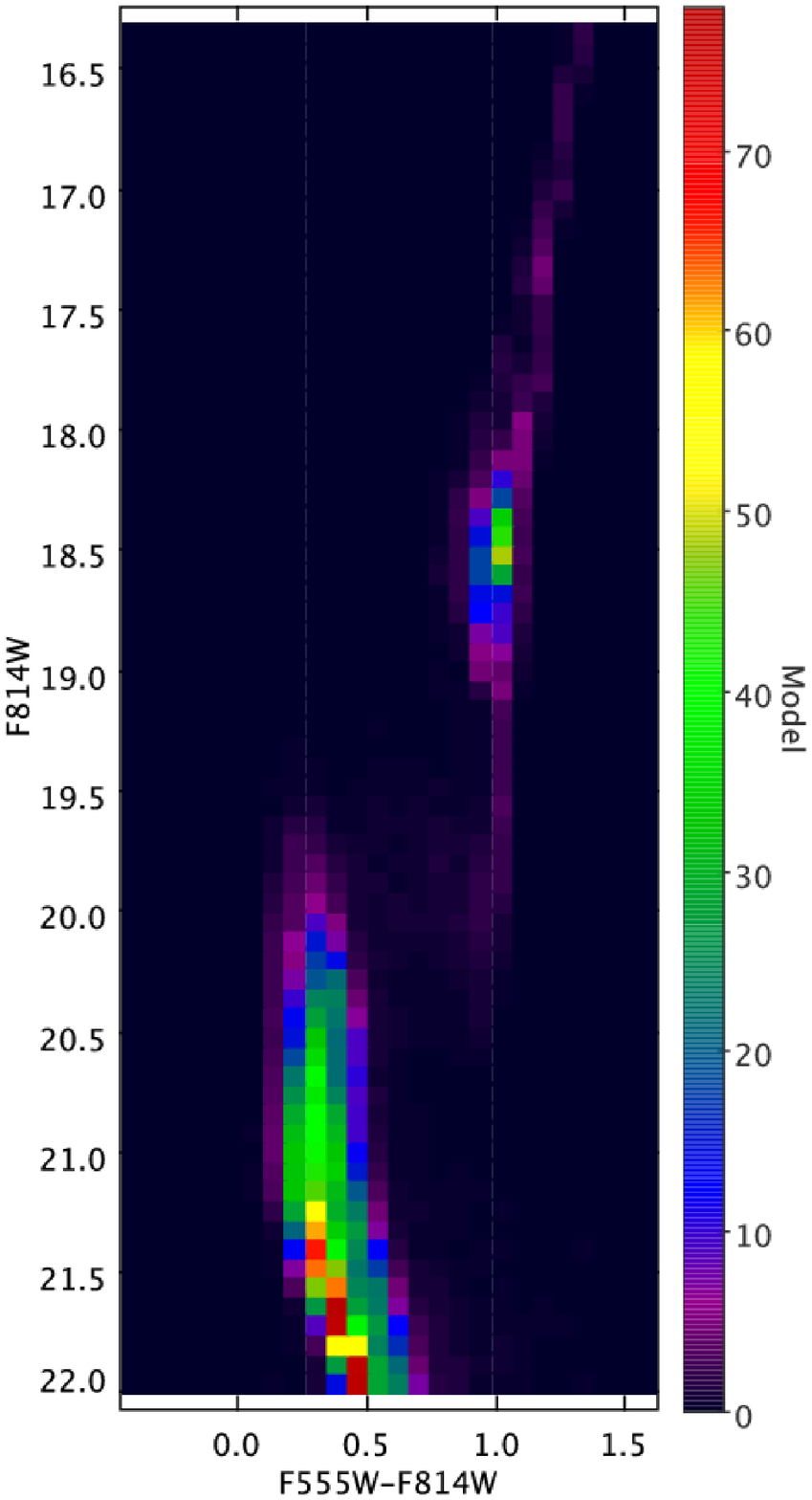}}
\end{minipage}
\begin{minipage}{0.24\textwidth}
\resizebox{\hsize}{!}{\includegraphics{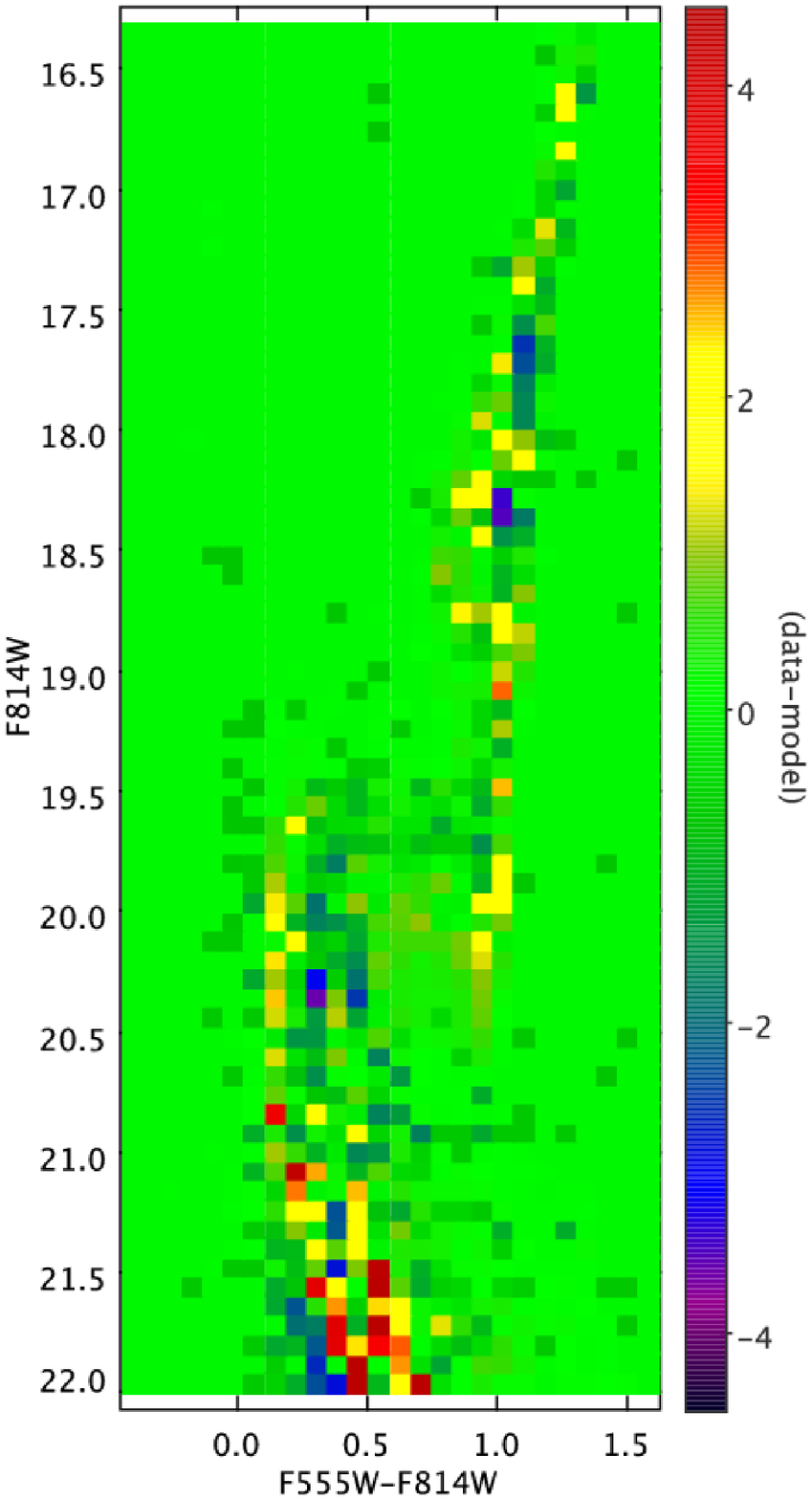}}
\end{minipage}
\begin{minipage}{0.24\textwidth}
\resizebox{\hsize}{!}{\includegraphics{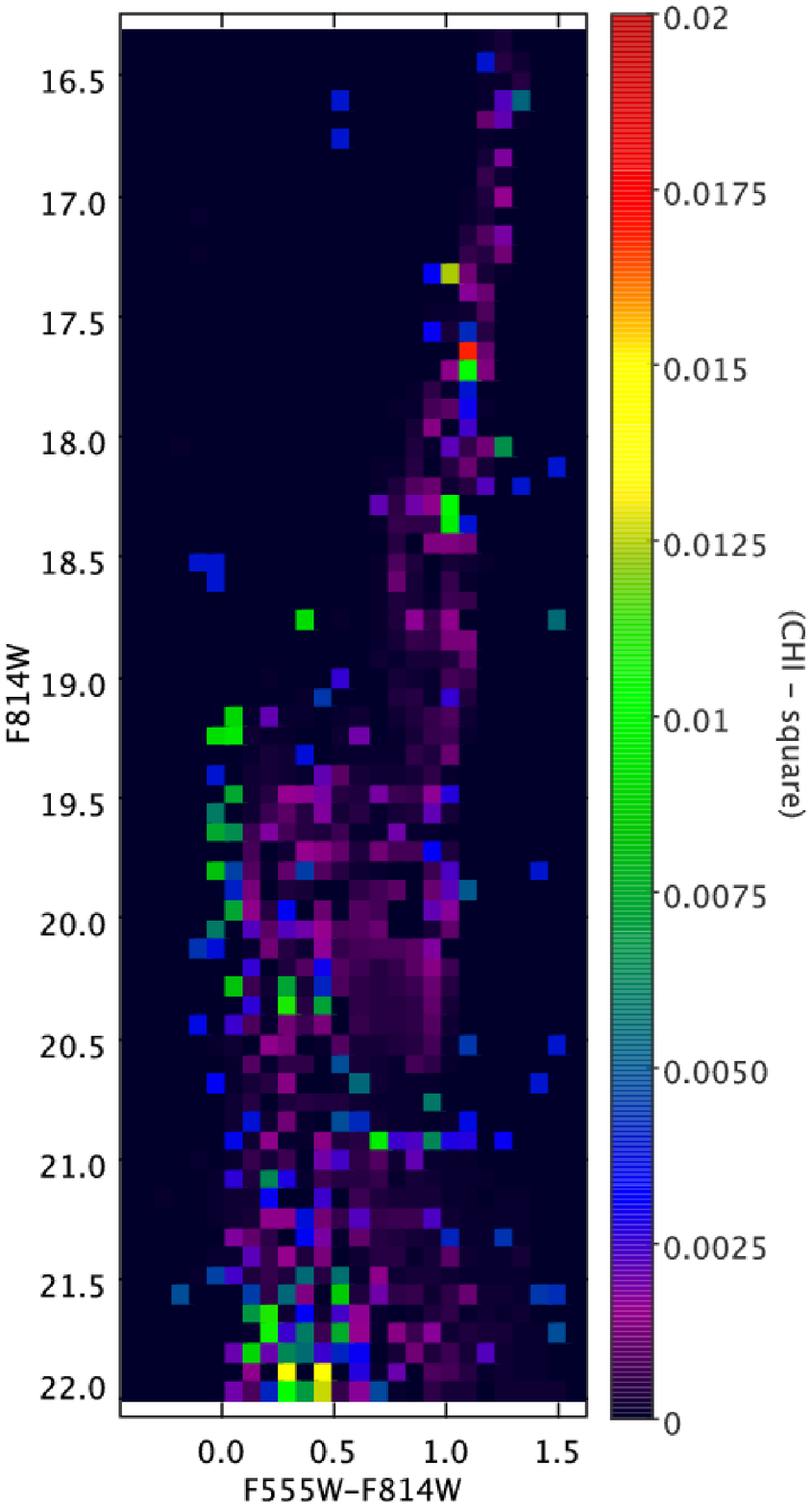}}
\end{minipage}
\caption{In the Hess diagram, we show the data (left), the solution
found by StarFISH (middle), the data-model difference and the $\chi^2$
map. The model is for $\feh=-0.88$, $\dm=18.83$, $\av=0.345$,
$\fb=0.18$.}
\label{fig_residuals}
\end{figure*}

\begin{figure*}
\begin{minipage}{0.48\textwidth}
\resizebox{\hsize}{!}{\includegraphics{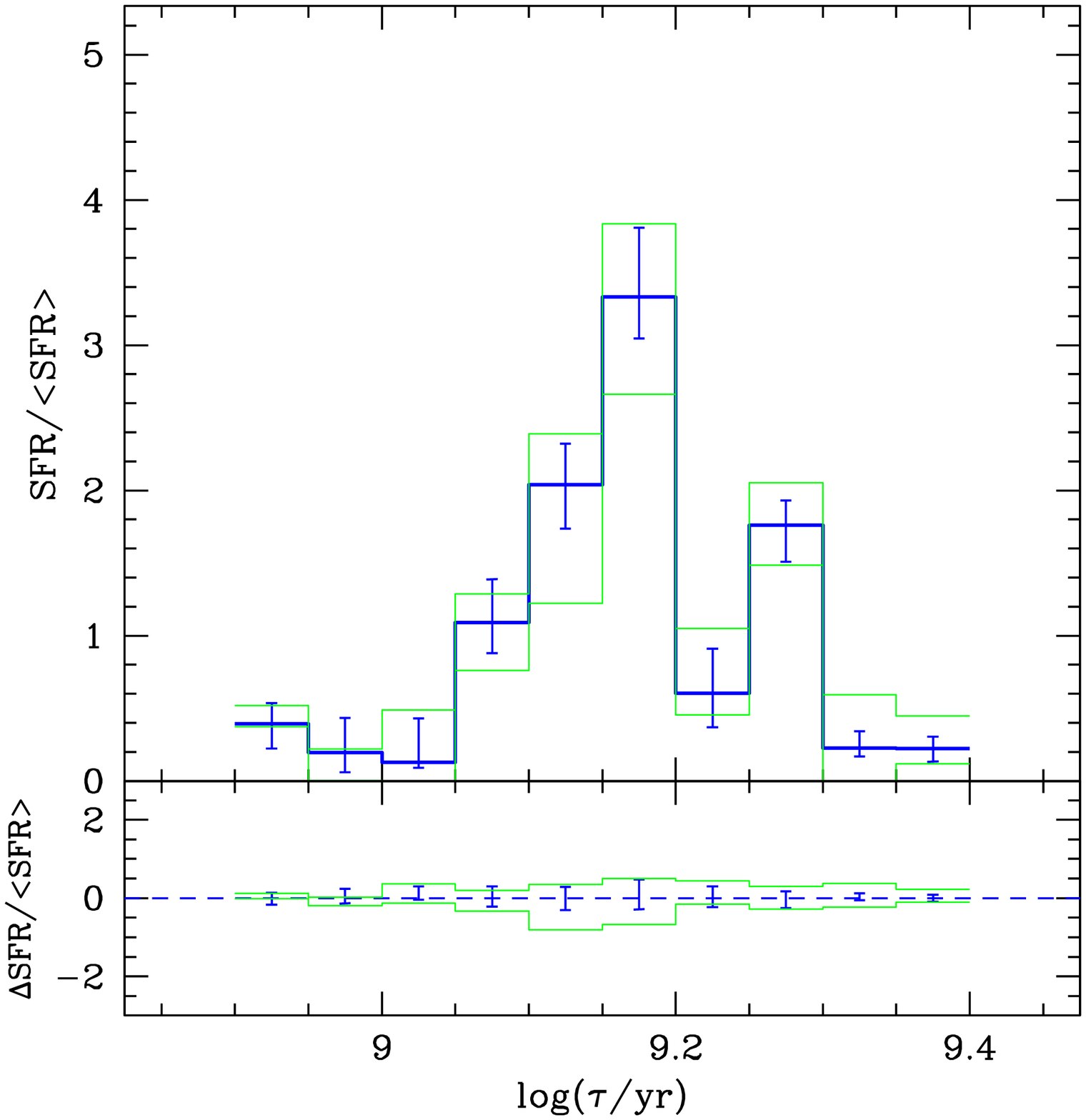}}
\end{minipage}
\begin{minipage}{0.48\textwidth}
\resizebox{\hsize}{!}{\includegraphics{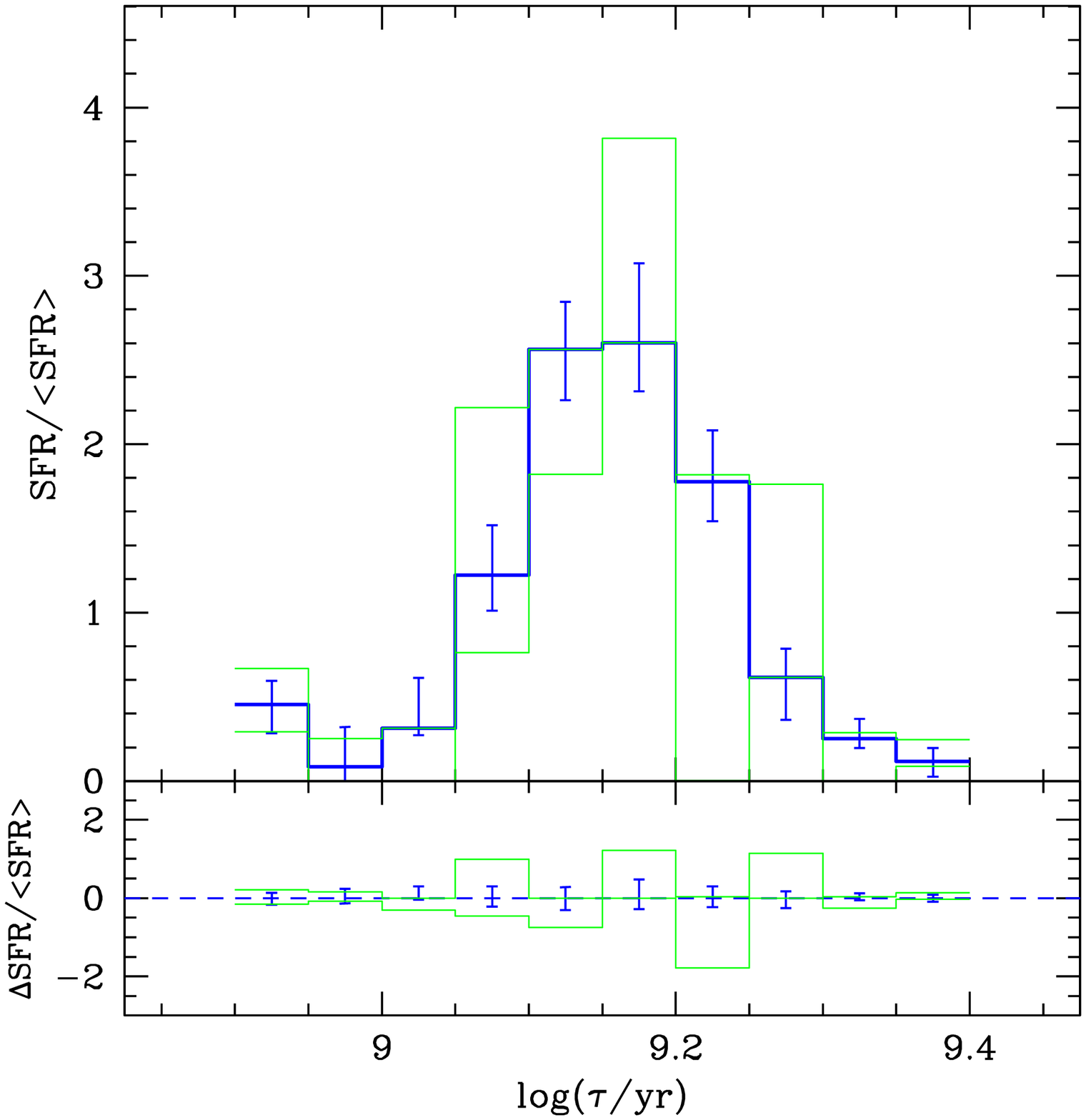}}
\end{minipage}
\caption{The SFR$(t)$ corresponding to the best fitting solution (blue 
histogram), for metallicities $\feh=-0.95$ (left) and $-0.81$
(right). The error bars indicate the random errors, whereas the green
histograms indicate the systematic errors (see
Sect.~\ref{sec_erroranalysis} for more details).}
\label{fig_sfh_extreme}
\end{figure*}

Among these many SFH-recovery experiments, the most interesting ones
are obviously those with the relative minima and the absolute minimum
$\chisqmin$ values. Fig.~\ref{fig_residuals} shows the $\chi^2$ and
residuals in the Hess diagram for the best fitting solution, which is
for $\dmo=18.83$, $\av=0.345$ and $\feh=-0.88$ and $\chi^{2}_{\rm
min}=0.933$. It is already evident, in this plot, that the fitted
solution is an excellent representation of the observed data, with
residuals uniformly distributed over the CMD. The only point of the
CMD in which there seems to be some increased residual is the bottom
part of the main sequence, where according to
Figs.~\ref{fig_completeness} and
\ref{fig_photerrors}, the completeness is significantly smaller than 
1 (although by just 10 to 20~\%), and photometric errors are higher.

Figure~\ref{fig_sfh_extreme} shows two examples of recovered SFR$(t)$ 
for best-fitting solutions obtained for different values of \feh. They are
qualitatively similar, with a clear presence of stars spanning ages 
from $\sim 1.2$~Gyr ($\log(t/{\rm{yr}})=9.08$) to $\sim~1.9$~Gyr 
($\log(t/{\rm{yr}})=9.28$).

\subsection{Evaluating the errors}
\label{sec_erroranalysis}

To evaluate the errors for all involved parameters the first step was
to find the correspondence between the $\chisqmin$ value for each
model and its significance (or confidence) level, $\alpha$.  This
correspondence was estimated simulating 100 synthetic CMDs generated
with a number of stars equal to the observed CMD, using the
best-fitting SFR$(t)$ and its parameters as the input for the
simulations. So, after recovering the SFH for this sample of synthetic
CMDs, it was possible to build the $\chisqmin$ distribution and to
establish the relation between the $\chisqmin$ difference above the
minimum and $\alpha$.

\begin{figure*}
\begin{minipage}{0.32\textwidth}
\resizebox{\hsize}{!}{\includegraphics{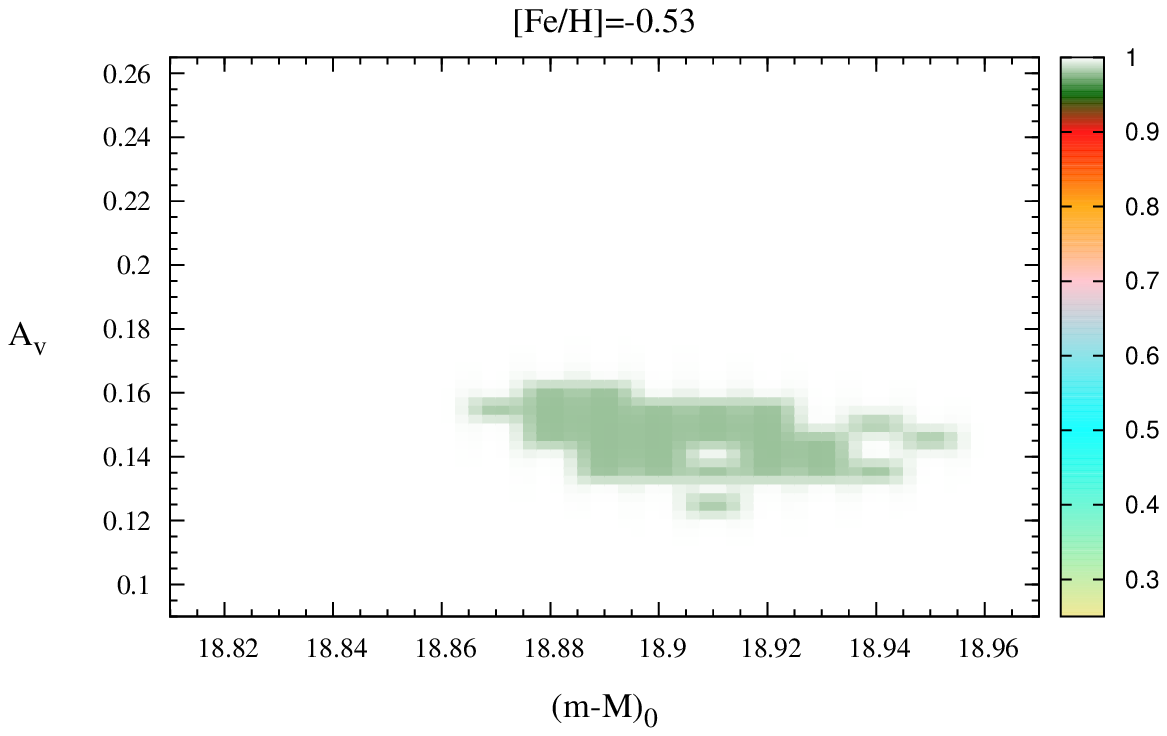}}
\end{minipage}
\begin{minipage}{0.32\textwidth}
\resizebox{\hsize}{!}{\includegraphics{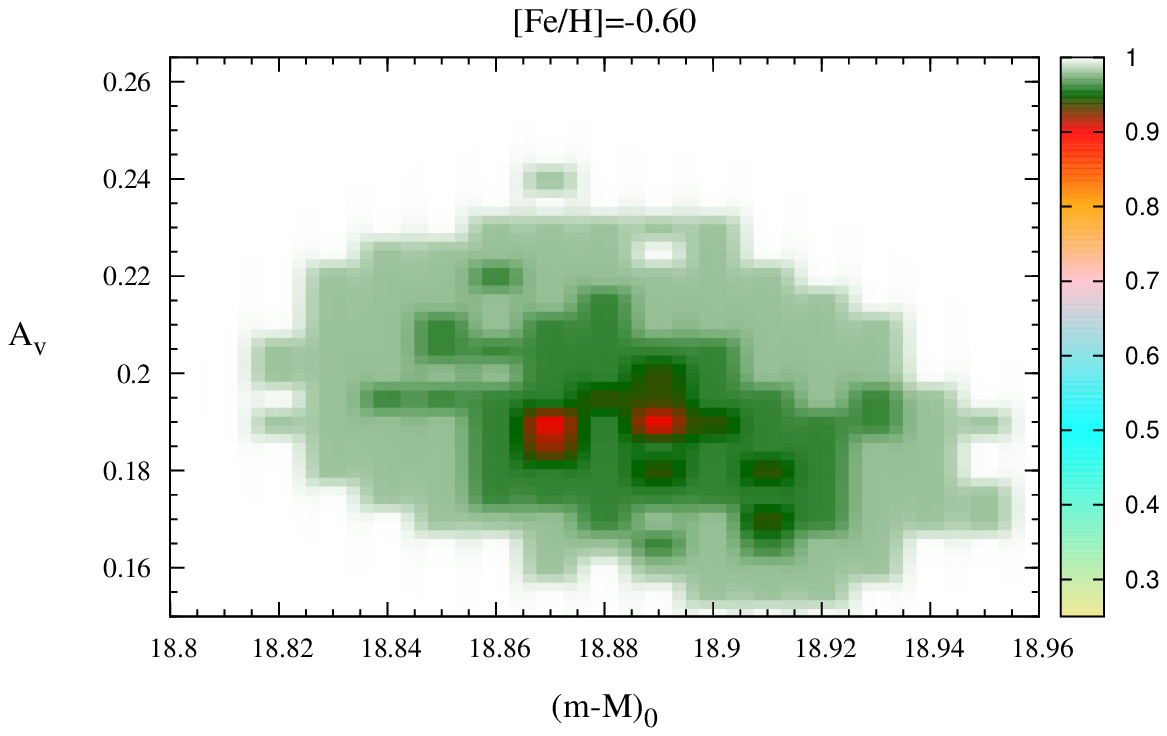}}
\end{minipage}
\begin{minipage}{0.32\textwidth}
\resizebox{\hsize}{!}{\includegraphics{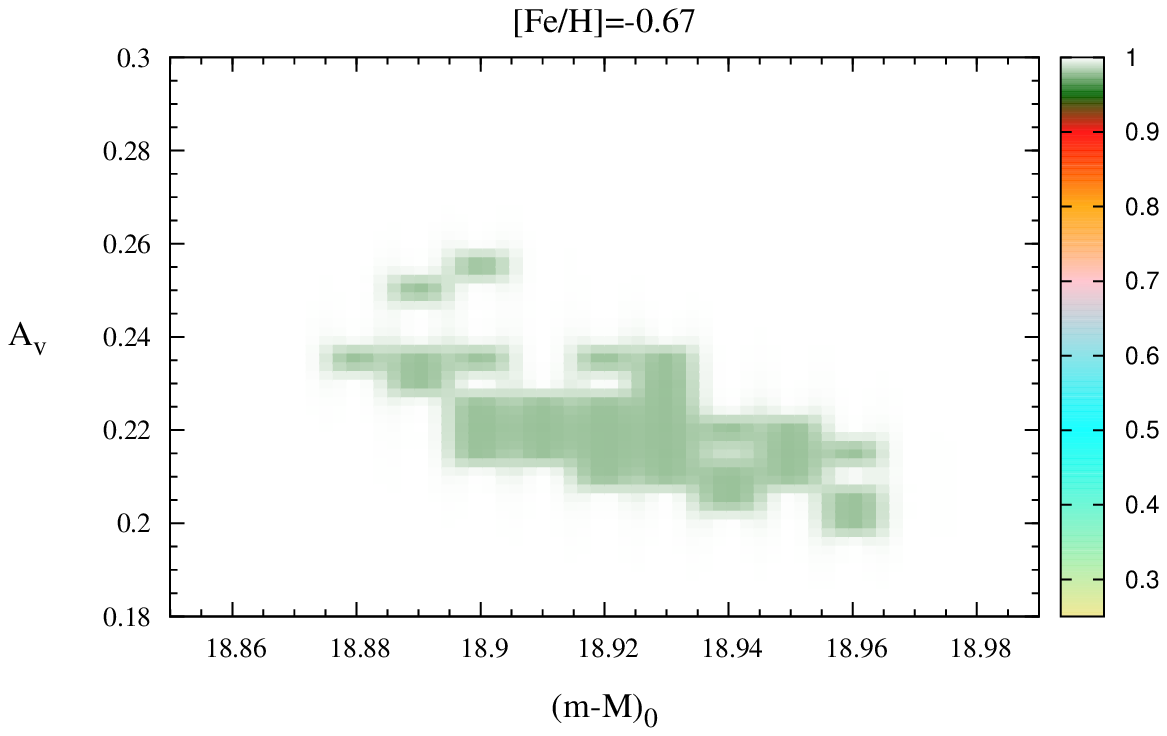}}
\end{minipage}
\begin{minipage}{0.32\textwidth}
\resizebox{\hsize}{!}{\includegraphics{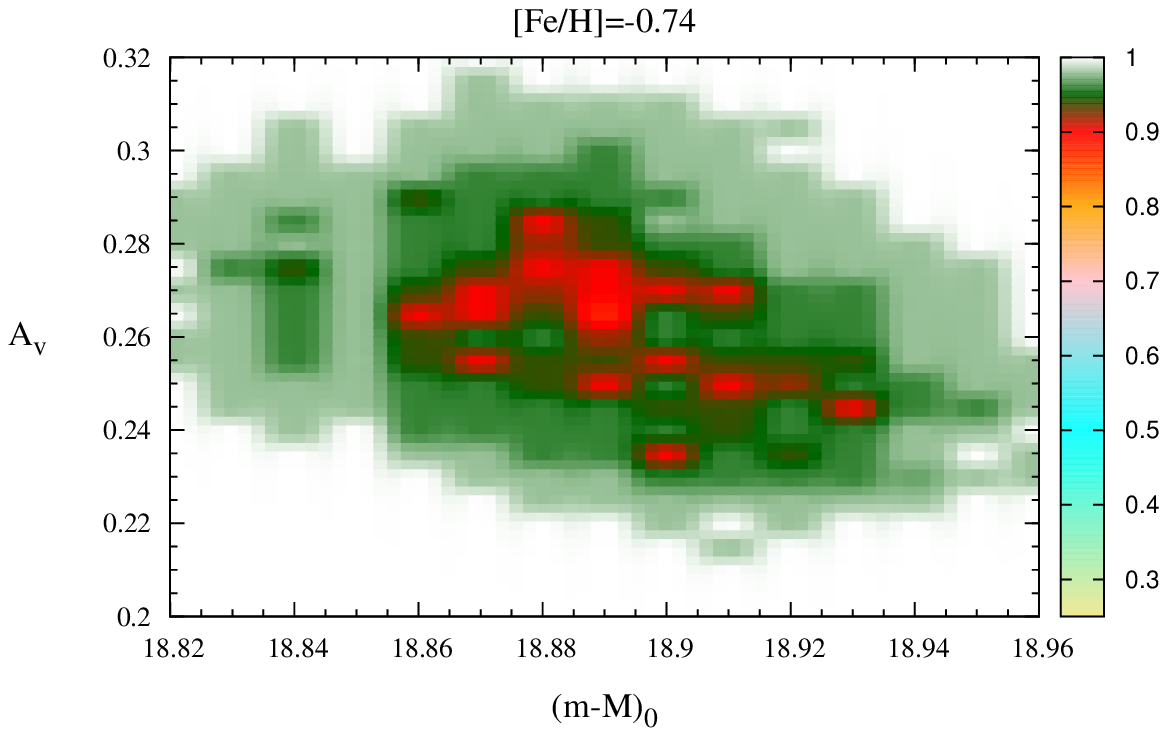}}
\end{minipage}
\begin{minipage}{0.32\textwidth}
\resizebox{\hsize}{!}{\includegraphics{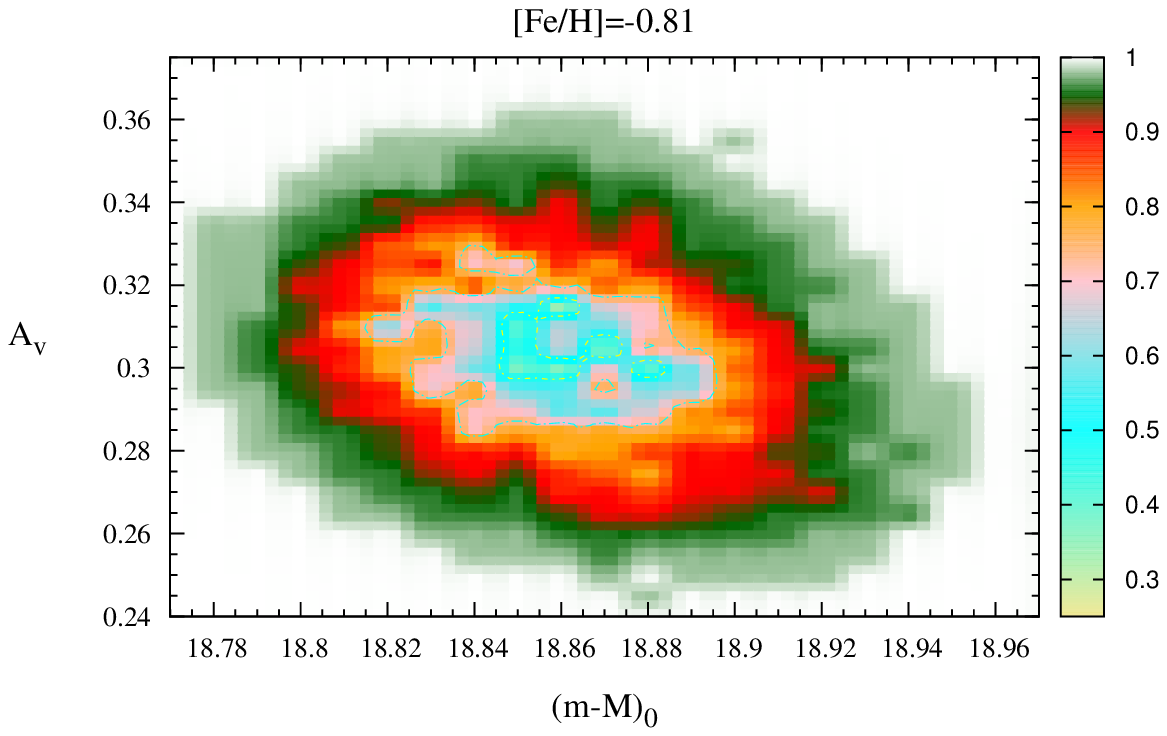}}
\end{minipage}
\begin{minipage}{0.32\textwidth}
\resizebox{\hsize}{!}{\includegraphics{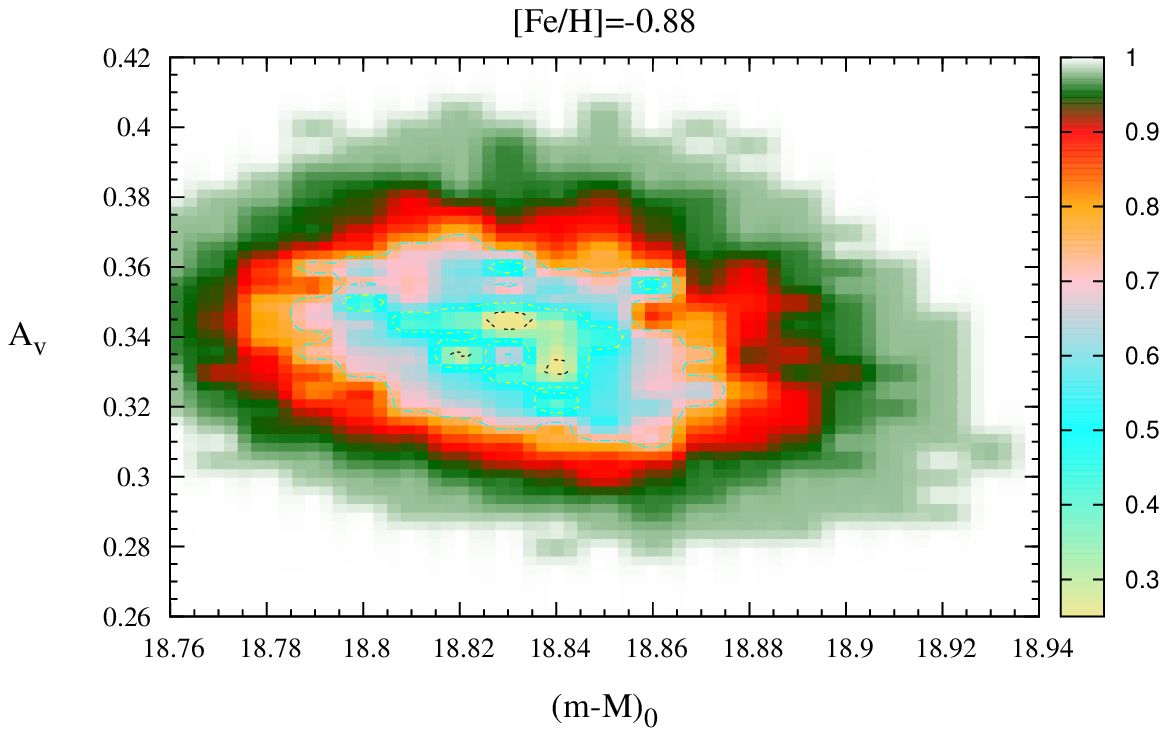}}
\end{minipage}
\begin{minipage}{0.32\textwidth}
\resizebox{\hsize}{!}{\includegraphics{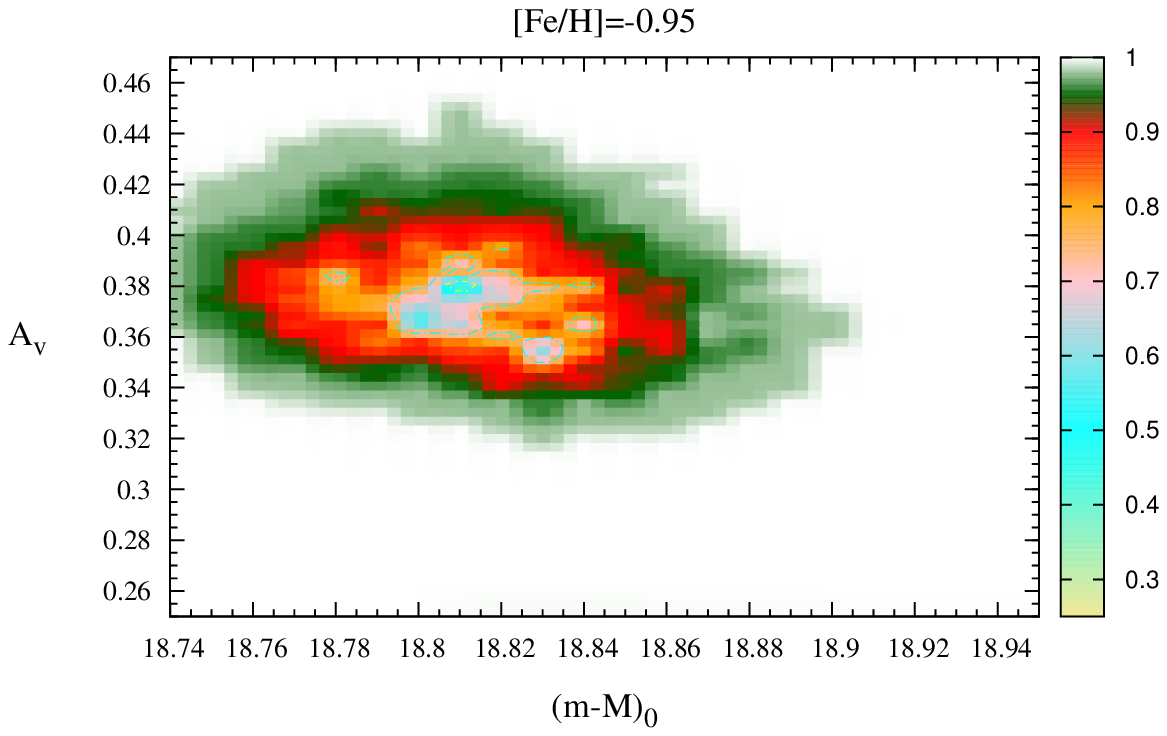}}
\end{minipage}
\begin{minipage}{0.32\textwidth}
\resizebox{\hsize}{!}{\includegraphics{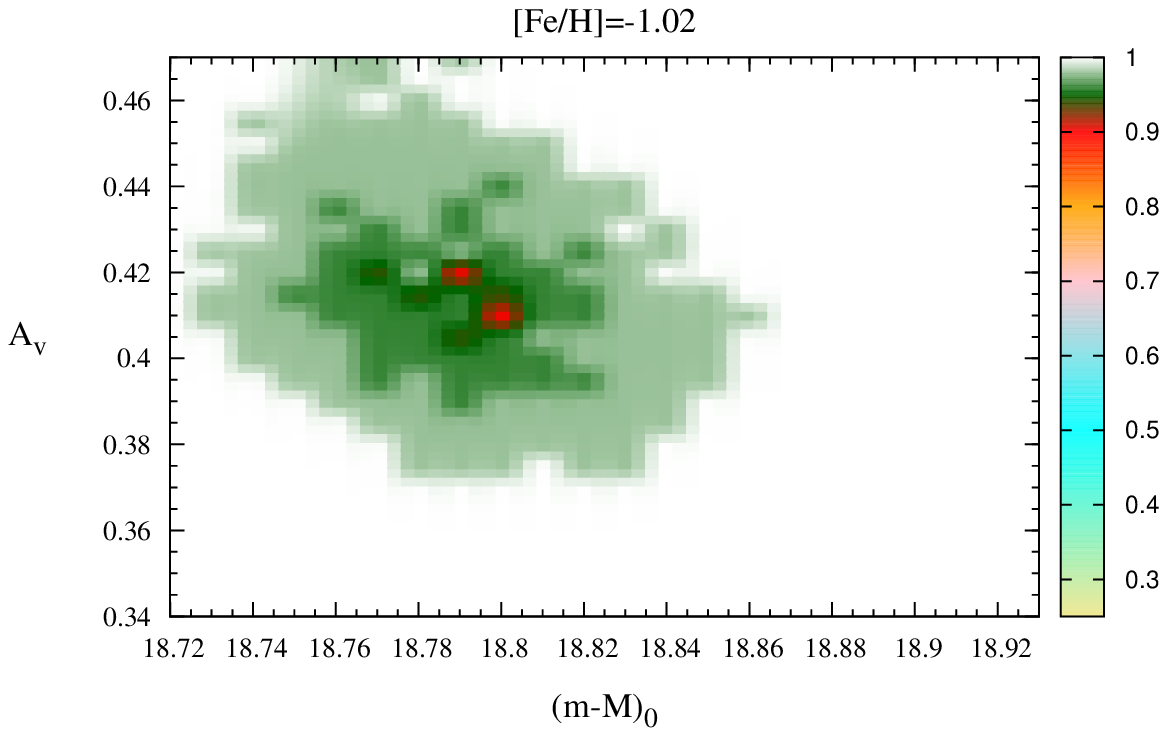}}
\end{minipage}
\begin{minipage}{0.32\textwidth}
\resizebox{\hsize}{!}{\includegraphics{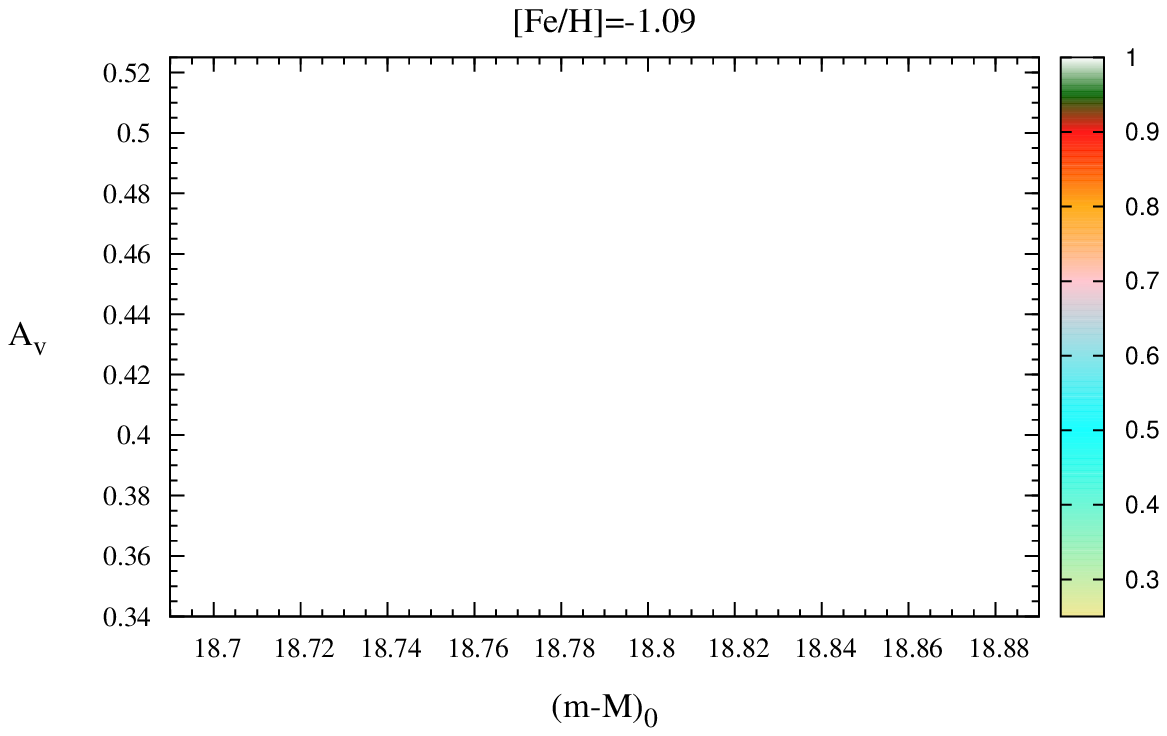}}
\end{minipage}
\caption{Significance level distributions for several \feh\ values.}
\label{prob_map}
\end{figure*}

Figure~\ref{prob_map} shows the significance levels for all the
solutions depicted in Fig.~\ref{chi2_map}. As one sees, for all \feh\
between $-0.74$ and $-0.95$ there are ample areas of the \av\ versus
\dmo\ diagrams with solutions within the 95~\% significance level. 
The best solutions, with $\chisqmin$ close to 0.95 and significance
levels of about 30~\%, are limited to the central regions of the
$\feh=-0.81$, $-0.88$, and $-0.95$ diagrams.

These maps can be used to estimate the mean values of the parameters
that characterize the best-fitting solutions, and their errors. These
are determined as the mean and standard deviation, weighted by
$1\!-\!\alpha$, inside the regions in which $\alpha<0.68$. The results
are:
\begin{eqnarray*}
\dmo &=& 18.84 \pm 0.04 \\
\av &=& 0.33 \pm 0.05 \\
\feh &=& -0.86 \pm 0.09 
\end{eqnarray*}
These numbers give an idea of the region of parameter space actually
covered by the best-fitting solutions.

\begin{figure}
\resizebox{\hsize}{!}{\includegraphics{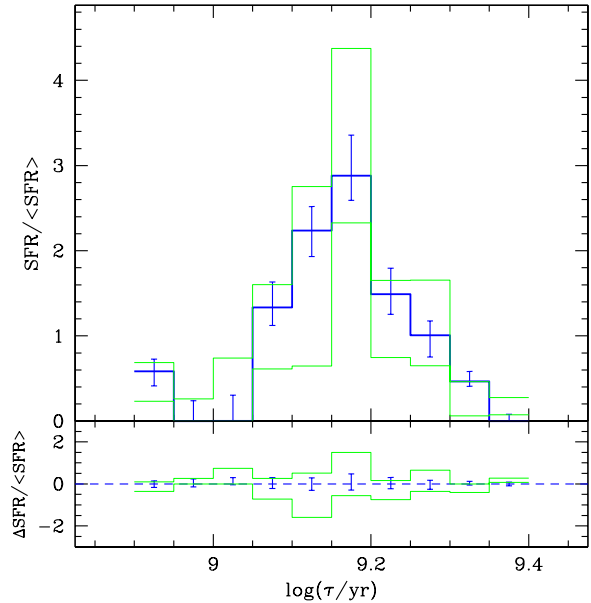}}
\caption{The overall best-fitting solution for the SFR$(t)$ of NGC~419. 
The random (error bars) and systematic errors (solid thin lines) are
also shown. }
\label{fig_sfh}
\end{figure}

Fig.~\ref{fig_sfh} presents the SFR$(t)$ for the overall best fitting
solution, obtained at $\feh=-0.88$, and corresponding to the Hess
diagrams of Fig.~\ref{fig_residuals}. Overplotted on the SFR$(t)$, we
have the random errors due to the statistical fluctuations (associated
to the number of stars); they were determined from the
r.m.s. dispersion in the recovered SFR$(t)$ for the 100 synthetic
CMDs.  On the other hand, the systematic errors in the SFR$(t)$ due to
the uncertainties in \dmo, \av\ and \feh, were determined using the
minimum and maximum SFR$(t)$ values for all models inside the 68~\%
significance level for all metallicities. 

\subsection{The role of the binary fraction}

\begin{figure}
\resizebox{\hsize}{!}{\includegraphics{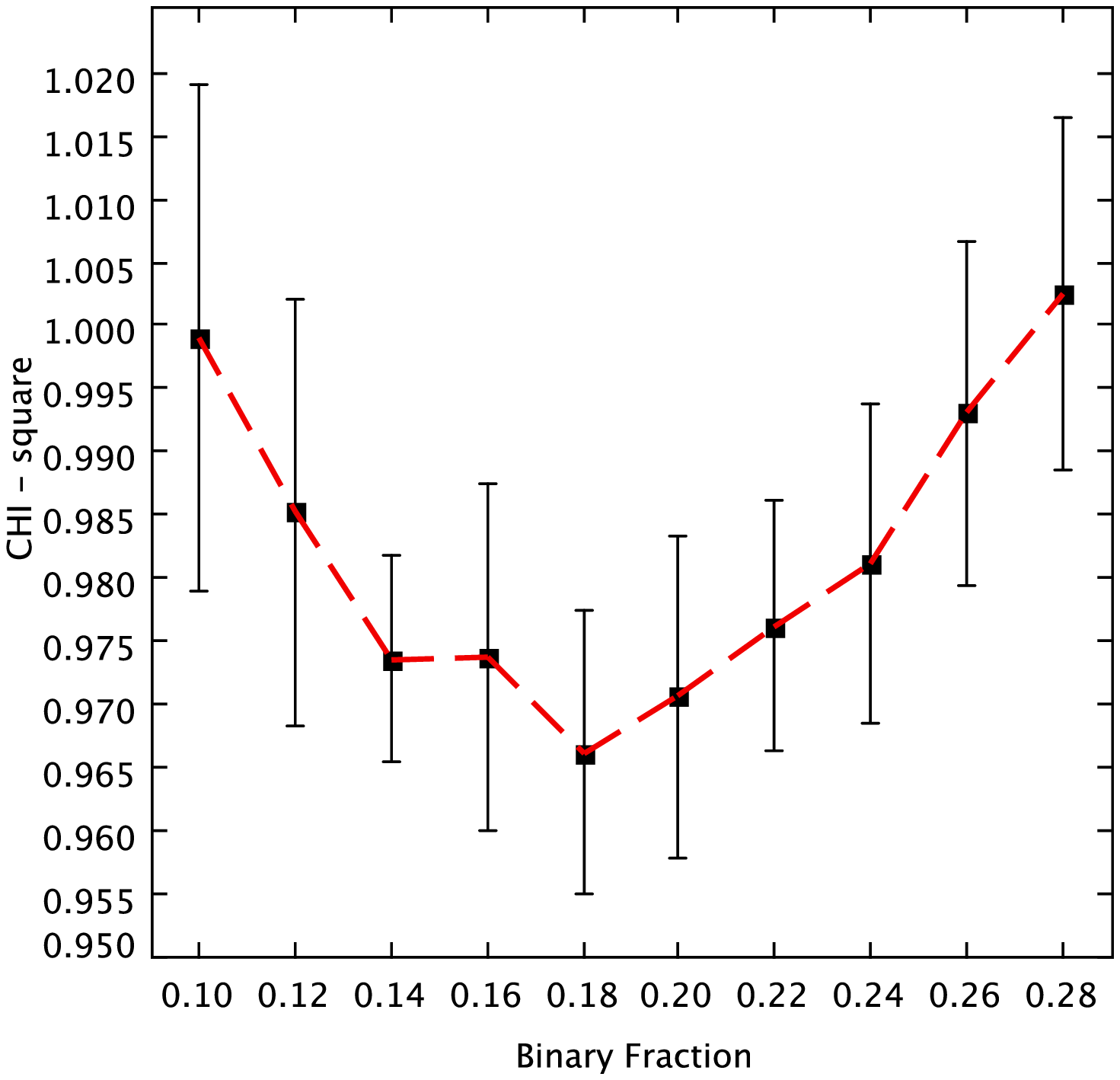}}
\caption{The average of the minimum $\chisqmin$ versus the binary 
fraction \fb, for a series of models of metallicity $\feh=-0.81$.}
\label{fig_fb}
\end{figure}
  
As anticipated previously, we have verified that the binary fraction
\fb\ plays a relatively minor role in determining the best-fitting
solution and the SFR$(t)$. Many simulations at varying \fb\ were tried
at several stages of our analysis, always supporting this conclusion,
and always pointing to a binary fraction close to
0.2. Fig.~\ref{fig_fb} aims to illustrate the sort of results one gets
for different \fb. To build the figure, we first define a large set of
SFH-solutions for which the CMD fitting was considered to be good:
namely the entire 68~\% significance level area for the metallicity
$\feh=-0.81$, in Fig.~\ref{prob_map}. For this region of the \av\
versus \dmo\ plane, the SFH-recovery tests are repeated for all \fb\
values between 0.10 and 0.28, at steps of 0.02. Then, we plot the
value of mean $\chisqmin$ and its r.m.s. dispersion as a function of
\fb, as shown in Fig.~\ref{fig_fb}. The minimum of mean $\chisqmin$ for
$\fb=0.18$ is evident. However, the dispersion of $\chisqmin$ values
is also significant, and indicates that a relatively wide range of
\fb, from say 0.12 to 0.24, would also produce acceptable results, if
compared to the $\fb=0.18$ case.

It is also worth noticing that for $\fb=0.18$ the 68~\% significance
level ($\alpha=0.68$) is about 0.038 above the minimum $\chisqmin$, as
determined in the previous section. All solutions presented in
Fig.~\ref{fig_fb} are inside this limit.  So, even if there is a
apparent minimum $\chisqmin$ around $\fb=0.18$, it is not
statistically significant.

\begin{figure}
\resizebox{\hsize}{!}{\includegraphics{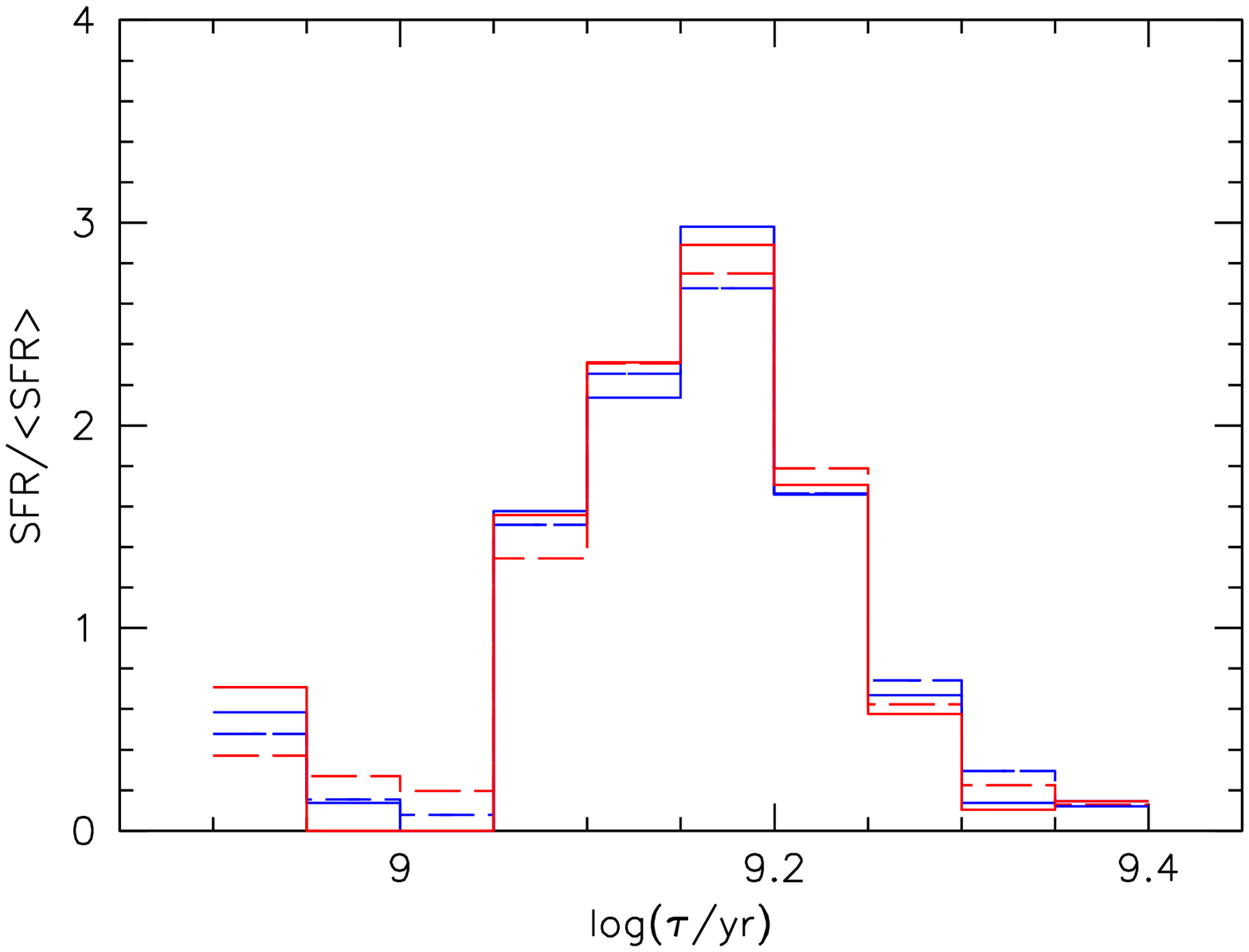}}
\caption{The SFR$(t)$ for a series of SFH solutions with
$\feh=-0.81$, $\dmo=18.86$, $\av=0.315$, and for different binary
fractions: $\fb=0.12$ (blue dashed line), 0.18 (red dashed), 0.24
(blue solid), and 0.28 (red solid).}
\label{fig_SFHfb}
\end{figure}

Finally, Fig.~\ref{fig_SFHfb} shows the recovered SFR$(t)$ for four
different \fb\ values over the $\fb=0.12$ to 0.28 interval. Not only
the SFR$(t)$ is seen to vary by amounts that are comparable to the
random errors in Fig.~\ref{fig_sfh}, but they also do it in a
non-systematic way: the mean age and shape of the SFR$(t)$ do not show
any significant trend with \fb.

\subsection{The role of the field contamination}

In this work, we have analysed the ACS/HRC photometry NGC~419 without
taking into account that a fraction of the observed stars is due to
the SMC field. Indeed, when first noticing the unusual structure of
the red clump in these data, \citet{Glatt_etal08} suggested the
fainter red clump could be due to the SMC field. The counter-argument
by \citet{Girardi_etal09}, however, has dramatically redimensioned
the possible role of the SMC field contamination in this case: after
measuring the density of red clump stars in the external parts of the
ACS/WFC images of the same cluster, at radii larger than $75\arcsec$
and for a total area of $2.47\times10^4$~arcsec$^2$, they find that
just 4.5 red clump stars from the SMC field are expected to be found
inside the 740~arcsec$^2$ area of the HRC images, whereas the total
observed number is of 388 (47 of them are in the secondary
clump). These numbers set the probability that the secondary clump in
NGC~419 is made of SMC field stars to less than $10^{-9}$
\citep{Girardi_etal09rio}. Moreover, they indicate that the field
contamination in the red clump area of the CMD is of just $\sim1$~\%.

In the context of the SFH-recovery work, some additional numbers worth
of consideration are the following. The number of field main sequence
stars in the magnitude interval $20<{\rm F814W}<22$, for the same
$2.47\times10^4$~arcsec$^2$ area in the outskirts of the WFC images,
is 2640, whereas the observed number in the 740~arcsec$^2$ of HRC is
2395. This magnitude interval is just barely affected by
incompleteness, and contains most of the observed stars. Therefore, we
can estimate that just $\sim3.3$~\% of the stars used to study the SFH
of NGC~419 are due to the SMC field.  This fraction is similar to the
Poisson noise in the total number of HRC stars, indicating that our
SFH results cannot be affected in a significant way.

\subsection{The age--metallicity relation}

With star formation lasting for 700~Myr, one may wonder whether
NGC~419 would not have enriched itself with metals ejected from its
oldest stellar populations. Although we believe that the present
photometric data is not enough to clarify this question, we performed
an additional SFH-recovery exercise, in which the cluster \feh\ is not
fixed, but can take four different values at each age: $\feh=-0.95$,
$-0.90$, $-0.85$, $-0.80$. In practice, this is obtained by running the
SFH-recovery method with a larger library of partial models, with 40
components (10 age bins times 4 metallicity bins). 

\begin{figure}
\resizebox{\hsize}{!}{\includegraphics{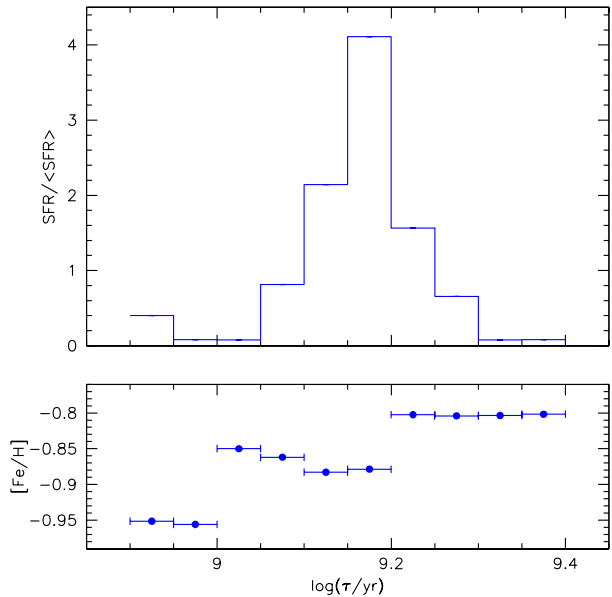}}
\caption{The SFR$(t)$ and AMR obtained when the \feh\ is not 
constrained to a single value, but is let to vary between 4 different
values in the $-0.95\le\feh\le-0.80$ interval. The upper panel
shows the \feh-added SFR$(t)$, whereas the bottom panel show the
SFR-averaged \feh$(t)$.}
\label{fig_SFHfeh}
\end{figure}

This larger set of partial models provides more freedom to the
StarFISH algorithm, which converges to an absolute best-fitting
solution with a \chisqmin\ of just 0.84, for $\av=0.33$ and
$\dmo=18.83$. The SFH results are illustrated in
Fig.~\ref{fig_SFHfeh}: the upper panel shows the SFR$(t)$ obtained by
adding the SFR values obtained for all metallicities at each age bin,
whereas the bottom panel shows the AMR as obtained by a SFR-weighted
average of the four different \feh\ values.

This SFR$(t)$ is actually very similar to the one obtained form the
best-fitting solution with $\feh=-0.88$ (Fig.~\ref{fig_sfh}). In the
age interval with significant star formation, which goes from 1.2 to
1.9~Gyr and comprises 5 age bins, the three youngest age bins are
found to have a metallicity almost coincident with the $\feh=-0.88$
one, whereas the two oldest ones are found to be slightly more
metal-rich, with $-0.80$~dex. It is not clear whether this \feh\
variation can be significant, since it is comparable to the estimated
error of $0.09$~dex in the \feh\ of the best-fitting solutions
(Sect.~\ref{sec_erroranalysis}). We conclude that our analysis does
not provide evidence for self-enrichment taking place in NGC~419.

\section{Conclusions}
\label{conclu}

In this paper, we have derived the SFH of the SMC star cluster
NGC~419, via the classical method of CMD reconstruction. This analysis
implicitly assumes that NGC~419 is formed by a sum of single-burst
stellar populations (or ``partial models''), each one being
characterized by a well-defined and narrow MSTO. The only effects that
blur these partial models in the CMDs are the sequences of binaries
(which effect however is far from dramatic) and the photometric
errors. With this kind of approach, the broad MMSTO observed in
NGC~419 naturally translate into a continued star formation
history. We find a SFR$(t)$ lasting for 700~Myr (from 1.2 to 1.9 Gyr,
see Fig.~\ref{fig_sfh}), which is quite a long period, probably at the
upper limit of all values already estimated for star clusters with
MMSTOs \citep{Bertelli_etal03, Baume_etal07, Mackey_BrobyNielsen2007,
Mackey_etal08, Milone_etal08, Goudfrooij_etal09}. Our error analysis
leaves practically no room for the MMSTOs in this cluster being caused
by a single episode of star formation.

It is also remarkable that the SFR$(t)$ we derive presents a
pronounced maximum at the middle of the star formation interval, at an
age of 1.5~Gyr (Fig.~\ref{fig_sfh}). This is somewhat unexpected, in
the context of the few suggested scenarios for the appearance of
prolonged star formation in LMC star clusters \citep[see][for a
comprehensive discussion of them]{Goudfrooij_etal09}. Either the
merging of two star clusters \citep{Mackey_BrobyNielsen2007}, or a
second period of star formation driven by the merging with a giant
molecular cloud \citep{Bekki_Mackey09}, would lead to strongly bimodal
distributions of cluster ages, which we do not find in NGC~419. Only
in the case of the best model with $\feh=-0.95$, there is an
indication of two different peaks in the SFR$(t)$ (see
Fig.~\ref{fig_sfh_extreme}), which however are neither separated nor
followed by periods of null SFR$(t)$. Also the trapping of field stars
by the forming star cluster \citep{PflammAltenburg_Kroupa07} would
lead to different results, with the major peak of star formation being
found at the youngest ages. On the other hand, our findings seem to be
more in line with \citet{Goudfrooij_etal09}'s conclusions, based on
the quite continuous distribution of stars across the MMSTO region of
the LMC cluster NGC~1846. They suggest a scenario in which the star
cluster continues to form stars in its center out of the ejecta of
stars from previous generations. In our case, however, this process
would have to proceed for a significantly more extended period of time
than for NGC~1846, and peak -- somewhat against the most naive
expectations -- at the middle of the star formation period. Also, this
latter scenario might imply some amount of self-enrichment in this
cluster, whereas our method instead is compatible either with a
constant metallicity, or with some amount of metal dillution. Needless
to say, the present observational data is not clear enough to provide
unambiguous indications about the formation scenario of such clusters.

Another basic result of our analysis is that the hypothesis of
continued SFH, together with current stellar evolutionary models and a
modest fraction of binaries, produces {\em excellent fits to the CMD
of NGC~419}, with $\chisqmin$ as small as 0.93 -- or even 0.84 if
the \feh\ is let to vary during the SFH-recovery. We have translated
this result into quantitative estimates for the random and systematic
errors of the derived SFR$(t)$. It is obvious that many combinations
of cluster parameters produce acceptable solutions, with significance
levels smaller than 95~\%. However, the really good solutions --
i.e. those with significance levels better then 68~\% -- cover a
narrow region of the parameter space, comprising less than 0.1 mag in
both \dmo\ and \av, and about 0.1~dex in \feh.

Despite our success in fitting the CMD of NGC~419 as a sum of partial
models, this success does not tell us about the reliability of
alternative hypotheses for the origin of MMSTOs. In particular, we
cannot conclude anything about \citet{Bastian_deMink09}'s hypothesis
based on the colour spread of coeval stars with different rotation
rates, since its capability of providing a {\em quantitative}
description of the data has not yet been tested. We urge that this
test should be performed, in order to shed light into this problem.

However, we call attention to another question that has to be
properly addressed: Is \citet{Bastian_deMink09}'s rotation hypothesis
able to produce, in addition to the broad turn-off of Magellanic Cloud
star clusters, also the dual red clump observed in NGC~419
\citep[see][]{Girardi_etal09}\,? In our models with prolonged star
formation, the 700~Myr spread in age translates into a
$\sim0.26$~\Msun\ spread in turn-off masses, which in turn implies a
small spread in the mass of H-exhausted cores as stars leave the main
sequence. This small spread -- of the order of just $\sim0.01$~\Msun\
-- is enough to cause the appearance of a dual red clump in this
cluster, as thoroughly discussed in \citet{Girardi_etal09,
Girardi_etal09rio}. Such a feature is then {\em naturally present} in
our best-fitting solutions, as can be appreciated in
Fig.~\ref{fig_residuals}. Can rotation do the same, producing spreads
of a compatible magnitude both in the main sequence and in the red
clump\,? Moreover, can this be achieved with a reasonable and simple
enough distribution of rotation velocities\,? At first sight, this
seems very unlikely to us.

\section*{Acknowledgments}
We thank the anonymous referee for the useful suggestions.
The data presented in this paper were obtained from the Multimission
Archive at the Space Telescope Science Institute (MAST). STScI is
operated by the Association of Universities for Research in Astronomy,
Inc., under NASA contract NAS5-26555.  We thank the support from
INAF/PRIN07 CRA 1.06.10.03, contract ASI-INAF I/016/07/0, and the
Brazilian agencies CNPq and FAPESP.



\label{lastpage}

\end{document}